\documentclass[dvips,epsfig,12pt]{article}

\usepackage{epsfig}

\input epsf.tex
\newcommand{\unit}[1]{\; \mbox{#1}}

\newcommand{\lsim}{\raisebox{-0.13cm}{~\shortstack{$<$ \\[-0.07cm] $\sim$}}~}

\newcommand{\eeen}{\mbox{$\times 10^{16}$}}

\begin{document}

\begin{center}
{\Large \bf Supersymmetric dark matter in M31:\\ 
can one see neutralino annihilation with CELESTE?}
\begin{flushright}
October 7th, 2002 \\
preprint GAM-2002/03 \\
preprint LAPTH-937\\
preprint PM/02-03

\end{flushright}

\vspace{1.0 cm}

\end{center}
\begin{center}
{\large A. Falvard, E. Giraud, A. Jacholkowska, J. Lavalle,
E. Nuss, F. Piron, and M. Sapinski\footnote{on leave from
 Henryk Niewodniczanski Institute of Nuclear Physics in Cracow}}\\
{\it Groupe d'Astroparticules de Montpellier, UMR5139-UM2/IN2P3-CNRS,
Place Eug\`ene Bataillon - CC85, 34095 Montpellier, France}\\

{\large P. Salati and R. Taillet }\\
{\it Laboratoire de Physique Th\'eorique {\sc lapth},
Annecy--le--Vieux, 74941, France
and Universit\'e de Savoie, Chamb\'ery, 73011, France}\\

{\large K. Jedamzik and G. Moultaka }\\
{\it Laboratoire de Physique Math\'ematique et Th\'eorique,UMR5825-UM2/CNRS, 
 Place Eug\`ene Bataillon, 34095 Montpellier Cedex 5, France}
\end{center}




\begin{abstract}

It is widely believed that dark matter exists within galaxies and clusters
of galaxies. Under the assumption that this dark matter is composed of the 
lightest, stable supersymmetric particle, assumed to be the neutralino,
 the feasibility of 
its indirect detection via observations of a diffuse gamma-ray signal due to 
neutralino annihilations within M31 is examined. To this end, first the dark 
matter halo of the close spiral galaxy M31 is modeled from observations,
then the resultant gamma-ray flux is estimated within supersymmetric
 model configurations.
We conclude that under favorable conditions such as the rapid
accretion of neutralinos on the central black hole in M31 and/or
the presence of many clumps inside its halo with $r^{-3/2}$ inner
profiles, a neutralino annihilation gamma-ray signal is marginally
detectable by the ongoing collaboration CELESTE.


\end{abstract}

\newpage
\section{Introduction}

The existence of cosmic dark matter is required by a multitude of 
observations and arguments, such as the excessive peculiar velocities of 
galaxies within clusters of galaxies or gravitational arcs, 
indicating much deeper gravitational 
potentials within clusters than 
those inferred to be present from only the 
luminous matter content. Also an epoch of Big 
Bang nucleosynthesis predicts a fractional 
contribution of baryons to the critical density, $\Omega_b$, significantly 
smaller than the total $\Omega$ in form of clumpy matter  
inferred to exist from observations of large-scale 
galactic peculiar 
velocities (e.g., $\Omega\approx 0.3$). Establishing the nature of the 
dark matter is one of the outstanding problems and challenges in cosmology. 
Virtually all proposed candidates require physics beyond the standard model 
of particle physics. A candidate particularly well-suited to the
formation of the observed large-scale structure, is a weakly interacting 
particle with comparatively small velocities before the onset of structure 
formation (i.e. cold dark matter, herafter CDM). 

\noindent
Due to the small thermal velocity of CDM, fluctuations survive from the
early universe on all scales. CDM is a bottom up scenario in which structures
develop as small clumps collapse and undergo series of merging resulting in 
the hierarchical formation of massive dark matter halos. These halos are
the hosts of baryonic systems, from galaxies to clusters, which cooled
and condensed by dissipating energy. Cosmological N-body simulations describe
well this non-linear complex series of mergers, accretion events and violent
relaxation which are in general agreement with observations but are subject to 
some difficulties on galactic scales. An overview of these issues was recently 
presented by ~\cite{pri}. It remains to be seen if the discrepancies 
on small scales may 
be solved by the particulars of baryonic physics (e.g. star formation, 
feedback, galactic black holes), or if indeed, the paradigm of cold dark 
matter is being challenged. In contrast, the excellent performance of 
cold dark matter (with inital adiabatic, scale-invariant density 
perturbations) on large scales seems still undiminished.

\noindent
On the particle physics side, supersymmetric models are believed to provide
the most promising approach to physics beyond the standard model. 
Supersymmetry can cure in principle several conceptual shortcomings
of this model, although definitively compelling scenarios are still to come.
Of interest for us here is the prediction  of, as yet undetected, bosonic 
particles for each fermionic standard model particle, and vice versa,
expected to lie in the approximate mass range of $\sim 100 GeV$ - $1 TeV$ 
which are actively searched for at present and future colliders. 
In particular, the lightest supersymmetric particle (LSP),
under the assumption of its stability (R-parity conservation) or at least
quasi stability on the cosmological time scale, may serve as the dark matter
particle as long as it is neutral (i.e. a neutralino). Moreover, its
interactions are believed such that it would behave as CDM, and its relic 
abundance may be naturally of order $\Omega\sim 1$ within the minimal
supersymmetric extension of the standard model (MSSM), for review
see ~\cite{mssm}. This is of
course a very exciting possibility.  

\noindent
There is a large number of ongoing underground experiments attempting to detect
neutralinos within the solar system by "direct" scattering on nuclei.
An alternative approach for the detection of neutralinos is via their 
occasional annihilation in dark matter halos, and observation of the
resultant gamma-rays (and/or neutrinos). 
However, when it comes to 
quantitative predictions in relation to experimental detection, one has to
tackle with astrophysical uncertainties such as the halo modelling
of the astrophysical object under consideration, as well as with particle 
physics uncertainties related to our ignorance of the physics underlying
supersymmetry breaking.
The possible
detection of such a gamma-ray signal from the spiral galaxy M31 
(Andromeda), at a distance of 700kpc, is the subject of the present
paper. We will deal mainly with two complementary issues: {\sl (i)} 
we give different models of the dark halo component in addition to the disk and bulge 
components previously studied in the literature; 
{\sl (ii)} assuming this dark matter
is accounted for by the lightest neutralino of the MSSM, we analyse the
gamma-ray fluxes that can originate from the pair annihilation of these
particles, in terms of model assumptions such as minimal supergravity. 

\noindent
The paper is organized as follows: section {\bf 2} is 
devoted to M31 halo modelling. In section {\bf 3} we first recall briefly the
basic ingredients and tools we use for the study of the supersymmetric
dark matter signature and then present the main predictions.
 Section {\bf 4} explores the discovery potential of the CELESTE experiment.
 Section {\bf 5} is
devoted to a quantitative discussion of possible signal enhancement due to
clumpiness and  black hole presence in M31.


\section{Modeling the neutralino halo around M31}

\subsection{Rotation curve of M31 without dark halo}
The late-type Sb spiral galaxy M31 lies at a distance of 700 kpc
. The visible part mostly consists of a bulge and a disk.
The present analysis is based upon the investigation of the neutral 
hydrogen content of M31, and the model-independent derivation of
the velocity field, performed by Braun ~\cite{bra}.
The rotation curve, after correction for the ellipticity which the spiral 
exhibits in the inner 5 kpc, is well fitted with two mass components 
which are both traced by optical observations:
{\sl (i)} A bulge -- with total mass $7.8 \pm 0.5 \times 10^{10}$ 
M$_{\odot}$ -- and assuming a mass-to-light ratio of 
$\Upsilon_{bulge} = 6.5 \pm 0.4~\Upsilon_{B,\odot}$ where 
$\Upsilon_{B,\odot}$ is the mass-to-light ratio for the Sun.
 This value
may be compared to the limit of $3.7 \leq \Upsilon_{bulge} \leq 5.7$ that has
been derived from synthetic stellar models ~\cite{gui} assuming the same
bulge color and a population age lying in the range from 9.5 Gyr to 15.5
Gyr; 
{\sl (ii)} The disk with a mass of $1.22 \pm 0.05 \times 10^{11}$
M$_{\odot}$ in the inner 28 kpc.

\begin{figure}[!t]
\centerline{\epsfig{file=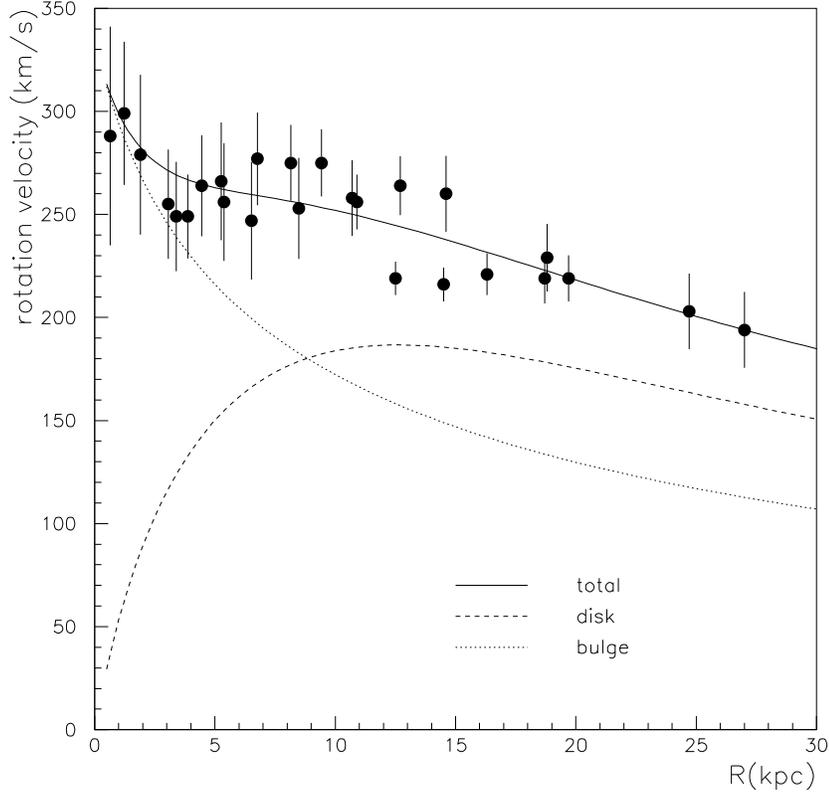,width=12cm}}
\caption{
the rotation velocity data points of the spiral Sb galaxy M31
are well fitted by the two-component model discussed in ~\cite{bra}.
The mass-to-light ratios -- in the blue band -- of the bulge and
the disk -- dotted curves -- are respectively $\Upsilon_{bulge} = 6.5 \pm 0.4
~\Upsilon_{B,\odot}$
and $\Upsilon_{disk} = 6.4 \pm 0.4~\Upsilon_{B,\odot}$. 
This leads to the global rotation
solid curve. Note that $\Upsilon_{disk}$ is too large to be consistent
with the young stars that populate the disk.}
\label{figure_RT1}
\end{figure}
%
\noindent
These contributions of the disk and the bulge components to the rotation
curve 
are plotted in Fig.~\ref{figure_RT1}. Note that this figure reproduces 
Fig.~8b of ~\cite{bra}.
Braun concludes that no dark halo is necessary to account for the velocity field 
inside M31. This result relies, nevertheless, on the crucial assumption that the
mass-to-light ratio of the disk is $\Upsilon_{disk} = 6.4 \pm 0.4
~\Upsilon_{B,\odot}$. Such
a large value does not agree with estimates based on the blue color of
the disk and on synthetic spectra of young stellar
populations which it contains ~\cite{gui}. 
The mass-to-light ratio $\Upsilon_{disk}$ of a purely stellar component
should actually not exceed 
$\sim$ $3.8~\Upsilon_{B,\odot}$. 
In addition, a disk as massive as that proposed by Braun
should generally be
unstable. We therefore feel that Braun has overestimated the importance
of the disk so that a halo around M31 is indeed a viable possibility.

\subsection{Rotation curve of M31 including a dark halo}

Since a disk and a bulge may not be enough to model the rotation
velocity of M31, we have assumed the presence of an additional
mass component in terms of a spherical halo whose mass density
profile is generically given by
\begin{equation}
\rho_{\chi}(r) \; = \; \rho_0 \,
\left( \frac{r_0}{r} \right)^\gamma \,
\left\{ {\displaystyle
\frac{r_0^\alpha + a^\alpha}{r^\alpha + a^\alpha}}
\right\}^\epsilon \;\; .
\label{neutralino_rho}
\end{equation}
A cored isothermal profile with core radius $a$
 corresponds to $\gamma = 0$, $\alpha = 2$ and
$\epsilon = 1$. A NFW
profile ~\cite{nav} is obtained with $\gamma = 1$, 
$\alpha = 1$, and $\epsilon =2$, whereas Moore's distribution ~\cite{bb6}
 is recovered if $\gamma = \epsilon = 3/2$ and $\alpha = 1$.
%
\begin{figure}[!t]
\centerline{
\epsfig{file=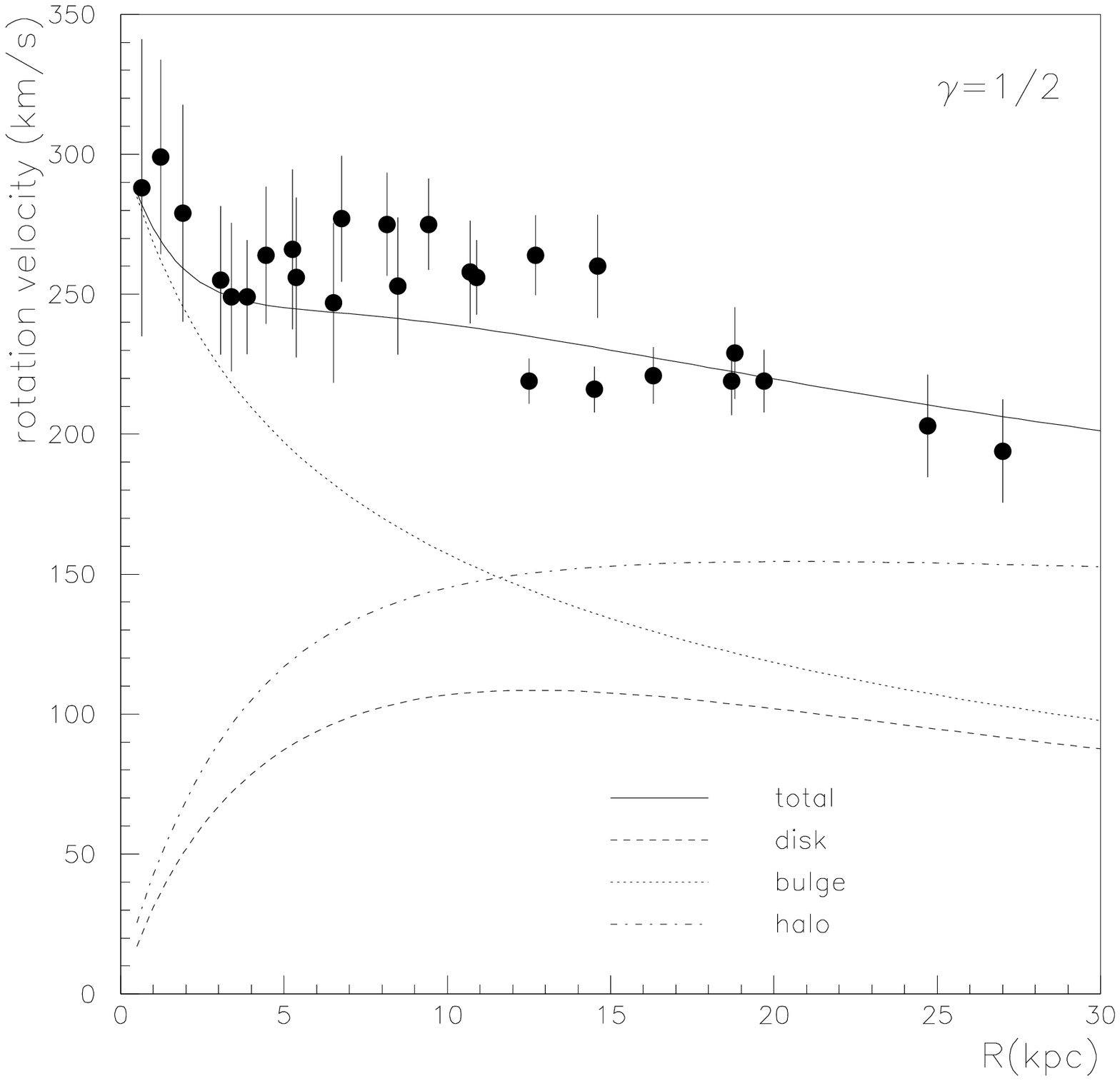,width=7.5cm}
\epsfig{file=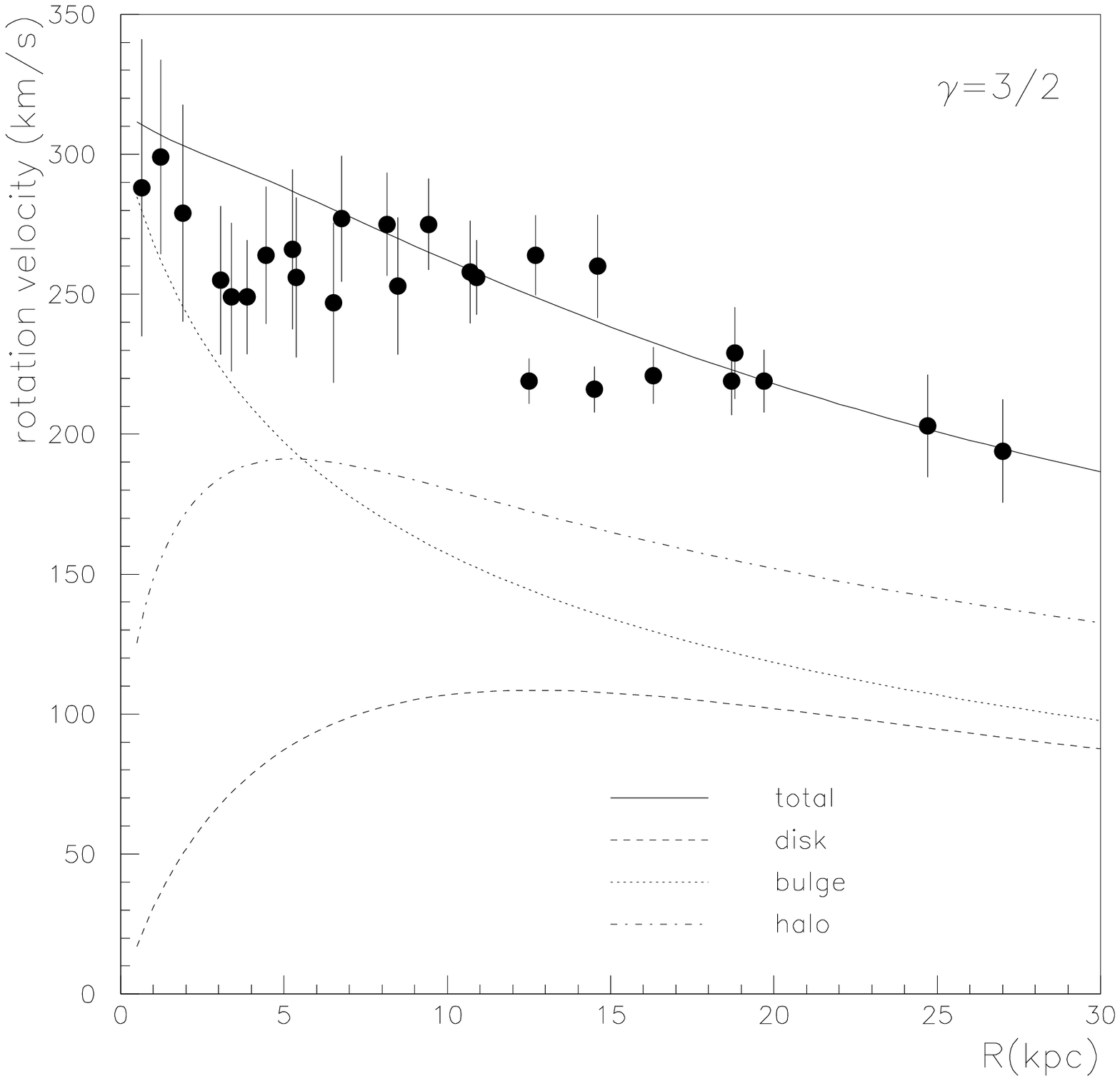,width=7.5cm}}
\caption{
a massive halo has been assumed in both plots by taking
small yet plausible values for the mass-to-light ratios
of the bulge -- $\Upsilon_{bulge} = 3.5~\Upsilon_{B,\odot}$
 -- and of the disk --
$\Upsilon_{disk} = 2.5~\Upsilon_{B,\odot}$. 
The left pannel corresponds to a value
of $\gamma = 1/2$ for the halo mass density singularity near the
center whereas the right panel features the case of a Moore's
profile with $\gamma = 3/2$. In both cases, the observations
are not well reproduced by the global rotation -- solid -- curves.}
\label{figure_RT2}
\end{figure}
%
For a given halo profile our mass models have two free parameters:
the mass-to-light ratio of the disk and that of the bulge.
 While there is ample
freedom on the relative importance of the disk and the halo, the
observations  set 
stringent constraints on the structure of the neutralino halo and actually 
favours a NFW ~\cite{nav} distribution. This is illustrated in Fig.~\ref{figure_RT2}
where the central singularity of the neutralino density has
been set equal to $\gamma = 1/2$  (left panel) and
$\gamma = 3/2$ (right panel). The
contributions of the bulge and the disk correspond to the dotted and
dashed curves, respectively, whereas the dashed-dotted lines denote the halo.
The global rotation -- solid -- curves fail to
match the observation points for medium range of R (a) and below 5 kpc
(b). We have therefore disregarded
these profiles in what follows and have concentrated on the NFW
($\gamma = 1$) case. This is not surprising insofar as a $1/r$ 
spherical profile leads to the same rotation curve as a disk with constant 
surface mass density.

\noindent
The neutralinos potentially concealed around M31 should annihilate and 
produce high-energy photons. The corresponding flux at Earth, $n_{\gamma}$
-- per unit of time, surface, and 
solid angle -- may be expressed as
\begin{equation}
\frac{d n_\gamma}{dt\,dS\,d\Omega}\; = \; \frac{1}{4 \pi} \,
{\displaystyle
\frac{\langle \sigma v \rangle \, N_{\gamma}}{m_{\chi}^{2}} } \,
{\displaystyle \int}_{\rm los} \rho_{\chi}^{2} \, ds \;\; 
= \; \frac{1}{4 \pi} \,
{\displaystyle
\frac{\langle \sigma v \rangle \, N_{\gamma}}{m_{\chi}^{2}} } \,
{\cal J}(R) \,
\label{gr_flux_1}
\end{equation}
where $m_{\chi}$ is the neutralino mass and
$\langle \sigma v \rangle \, N_{\gamma}$ denotes the thermally averaged annihilation rate yielding $N_{\gamma}$
gamma-rays in the final state. The astrophysical part
of the expression consists in the integral ${\cal J}$ along the
line of sight of the neutralino density squared $\rho_{\chi}^{2}$.
Assuming a spherical halo with radial extension $R_{\rm max}$ leads
to
\begin{equation}
{\cal J}(R) \; = \; 2 \,
{\displaystyle
\int_{0}^{\sqrt{R_{\rm max}^{2} - R^{2}}}} \,
\rho^{2} \bigg (\sqrt{s^2 + R^2}\bigg)\; ds \;\; ,
\label{gr_flux_2}
\end{equation}
for a direction with impact parameter $R$ off the centre of M31. Because the
density decreases steeply at large distances (see Eq.~\ref{neutralino_rho}),
our results are not sensitive to the actual value
of the radial cut-off $R_{\rm max}$.
The next step is the sum of the line of sight integral ${\cal J}$
over the solid angle subtended by the source
\begin{equation}
\Sigma \; = \; \int \, {\cal J}(R) \; d \Omega \;\; .
\end{equation}
We are interested in the number $I_{\gamma}$ of high-energy photons
-- collected per unit of time and surface -- that
originate from a circular region with angular radius $\theta_{\rm obs}$.
The previous expression simplifies into
\begin{equation}
\Sigma \; = \; 2 \pi \,
{\displaystyle \int_{0}^{\displaystyle \theta_{\rm obs}}}
\, {\cal J}(R) \; \sin \theta \; d \theta \;\; ,
\end{equation}
where $R / D = \tan \theta \simeq \theta$ and $D \sim 700$ kpc
is the distance to M31. Integrating relation~(\ref{gr_flux_1})
over the source leads to the gamma-ray signal
\begin{equation}
I_{\gamma} \; = \;
\left( 3.18 \times 10^{-13} \unit{photons} \unit{cm}^{-2}
\unit{s}^{-1} \right) \;
\left\{ {\displaystyle
\frac{\langle \sigma v \rangle \, N_{\gamma}}
{10^{-25} \unit{cm}^3 \unit{s}^{-1}}} \right\} \;
\left\{ {\displaystyle \frac{500 \unit{GeV}}{m_{\chi}}}
\right\}^{2} \; \Sigma_{19} \;\; ,
\label{I_gamma}
\end{equation}
where $\Sigma_{19}$ denotes $\Sigma$ expressed in units
of ${10^{19} \unit{GeV}^{2} \unit{cm}^{-5}}$.

%
\begin{figure}[!t]
\centerline{
\epsfig{file=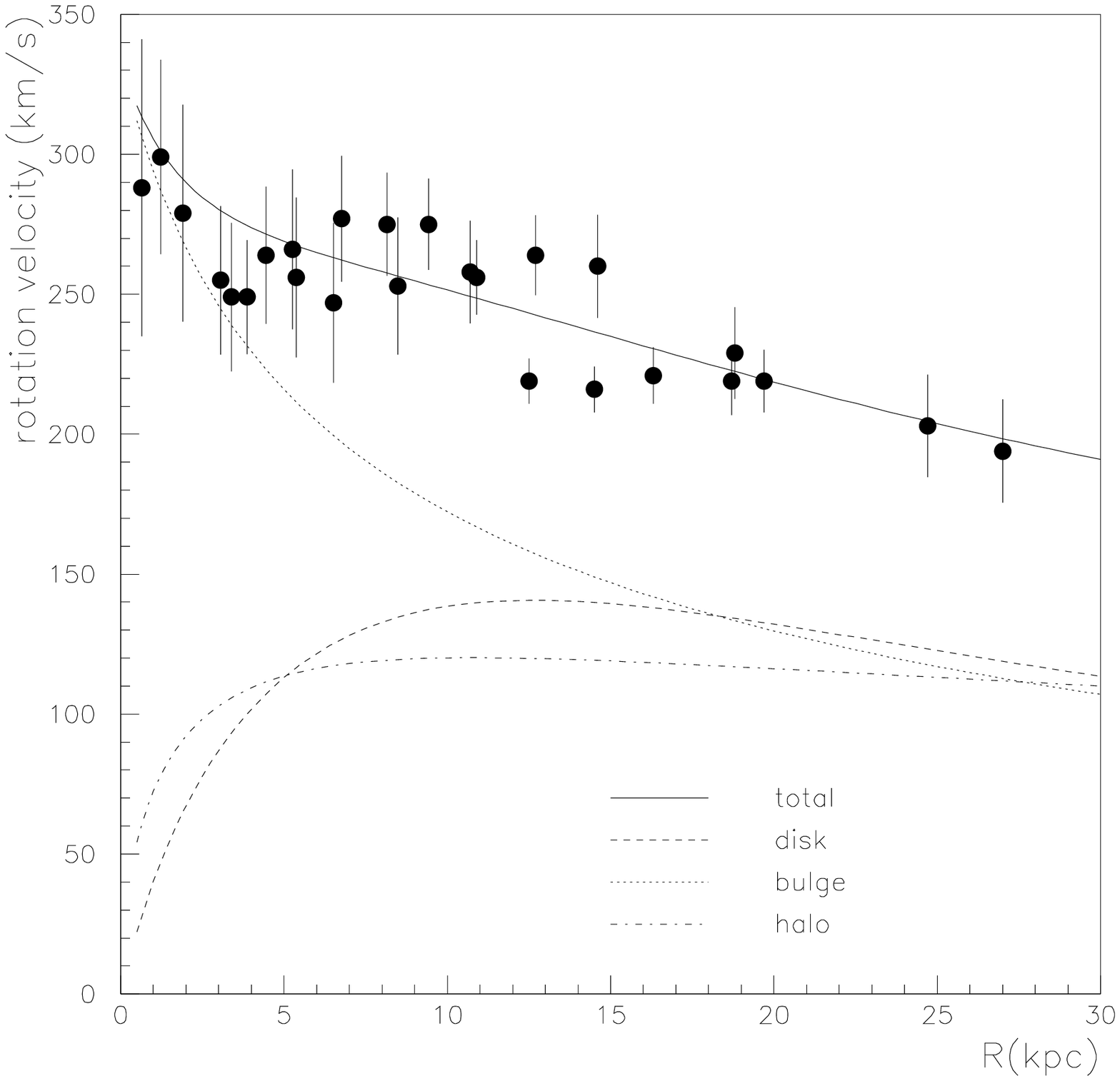,width=7.5cm}
\epsfig{file=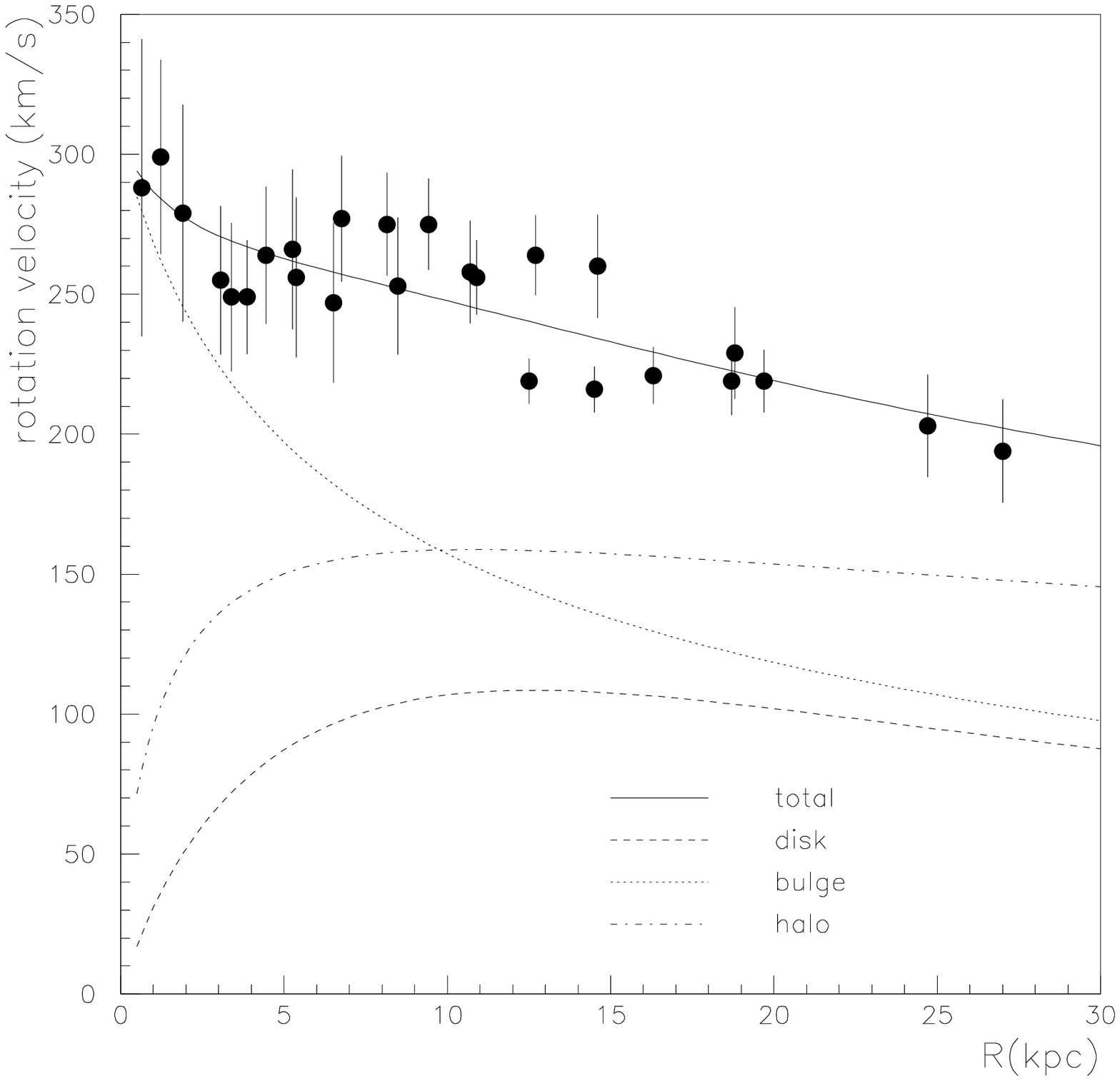,width=7.5cm}}
\caption{
a $\gamma = 1$ neutralino halo is added to the bulge and to the
disk of M31. In the left panel, an intermediate case is featured
with values for the mass-to-light ratios of $\Upsilon_{bulge} = 4.2
~\Upsilon_{B,\odot}$
and $\Upsilon_{disk} = 4.2~\Upsilon_{B,\odot}$. 
The right panel corresponds to a maximal
halo with the same mass-to-light ratios as in Fig.~\ref{figure_RT2}
 (see also Table~\ref{table_RT1}). The global -- solid -- rotation
curve is in good agreement with the data of ~\cite{bra}.}
\label{figure_RT3}
\end{figure}
%

\noindent
In order to estimate the gamma-ray emission of an hypothetical
neutralino halo around M31, we have considered
%
an intermediate NFW halo with bulge and disk mass-to-light ratios of 
$\Upsilon_{bulge} = 4.2~\Upsilon_{B,\odot}$ and $\Upsilon_{disk} = 4.2
~\Upsilon_{B,\odot}$ (left panel of 
Fig.~\ref{figure_RT3}). The parameter $\Sigma_{19}$ of
expression~(\ref{I_gamma}) reaches 1 in the inner 3.5 kpc which
corresponds to $5$ mrad, field of view of CELESTE and 1.2 up to 28 kpc
which is the M31 radius. 
A fiducial model for the neutralino
has been assumed with $m_{\chi} = 500$ GeV and annihilation cross
section multiplied by mean velocity and average number of $\gamma$
in final state: $\langle \sigma v \rangle \, N_{\gamma} \, = \,
{10^{-25} \unit{cm}^3 \unit{s}^{-1}}$.
The corresponding gamma-ray fluxes are respectively
%
$3.2 \times 10^{-13}$ and
$3.8 \times 10^{-13} \unit{photons} \unit{cm}^{-2} \unit{s}^{-1}$ as
presented in Table~\ref{table_RT1}. 
 In the right panel of
Fig.~\ref{figure_RT3}, a more massive halo has been considered.
The mass-to-light ratios of the bulge and of the disk are 
$\Upsilon_{bulge} = 3.5~\Upsilon_{B,\odot}$ and 
$\Upsilon_{disk} = 2.5~\Upsilon_{B,\odot}$, respectively. 
In this more favourable
situation for detection, the gamma-ray
emission is increased by a factor of 3 with respect to the
previous case.
Recent data at radial distance $ 15 \le r \le 30 $ kpc, employing different data
reduction analysis than that by Braun imply the presence of more mass at large radius than in
Braun's curve and a flatter mass distribution ~\cite{bb37}. Therefore
 one may well be close to the most favourable case.

%
%
%
\begin{table}[h!]
\[
\begin{array}{|cccccc|}
\hline
\hline
     \Upsilon_{bulge} & \Upsilon_{disk} & \Sigma_{19}(3.5 \unit{kpc}) &
     \Sigma_{19}(28 \unit{kpc}) & I_{\gamma}(3.5 \unit{kpc}) &
     I_{\gamma}(28 \unit{kpc}) \\
     \hline
     6.5 & 6.4 & 0 & 0 & 0 & 0 \\
     4.2 & 4.2 & 1  & 1.2  &
     3.2 \times 10^{-13} & 3.8 \times 10^{-13} \\
     3.5 & 2.5 & 3  & 3.7  &
     10.2 \times 10^{-13} & 11.8 \times 10^{-13} \\
\hline
\hline
\end{array}
\]
\caption{
three different models for M31 are featured in this table. When
the mass-to-light ratios of the bulge and of the disk are respectively
$\Upsilon_{bulge} = 6.5~\Upsilon_{B,\odot}$ and 
$\Upsilon_{disk} = 6.4~\Upsilon_{B,\odot}$, no halo is necessary
(Fig.~\ref{figure_RT1}). The intermediate halo (second line) 
corresponds to the left panel of Fig.~\ref{figure_RT3}, and the massive 
halo (third line) to the right panel of
Fig.~\ref{figure_RT3}. The line of sight integral $\Sigma_{19}(R)$ is expressed
in units of $10^{19}$$\unit{GeV}^{2} \unit{cm}^{-5}$ whereas the 
flux $I_{\gamma}$(R)
corresponds to the number of photons collected per $\unit{cm}^{2}$ and per
sec, for a circular region encompassing the inner 3.5 and 28 kpc, respectively
. The assumed neutralino mass is $m_{\chi} = 500$ GeV 
with an annihilation cross
section such that $\langle \sigma v \rangle \, N_{\gamma} \, = \,
{10^{-25} \unit{cm}^3 \unit{s}^{-1}}$.
}
\label{table_RT1}
\end{table}
\section{ Supersymmetric model predictions}

In the present section we concentrate on more specific particle
physics model predictions of the $\gamma$ - ray fluxes.
 The aim is to implement,
test and use a computational tool which allows to scan over various
observables related to supersymmetric dark matter.
This is achieved through an interfacing of two public codes,
DarkSUSY \cite{DSUSY} and SUSPECT \cite{Suspect}, which we dub 
hereafter DSS (DarkSUSY-SUSPECT). Significant features on both particle physics
and cosmology sides are thus combined within our approach. 
For instance, we will study the effects of universality versus non-universality
of the soft supersymmetry breaking parameters and/or the (necessary)
requirement of radiative electroweak symmetry breaking, both on
the cosmological relic density of the neutralino LSP as well as on the 
gamma-fluxes from LSP annihilation in the halo of M31 as modelled in the 
previous section.      

\subsection{MSSM parameterization}
We will focus mainly on the minimal supergravity scenario (mSUGRA) 
, where the soft breaking of supersymmetry occurs in a hidden sector, 
which communicates with the visible sector only via gravitational interactions,
and translates to the observable particle physics sector in the form
of soft masses and trilinear couplings among the scalar fields as well as
mass terms in the gaugino sector \cite{mSUGRA}. 
We make the usual simplifying assumption of
 common universal values of these parameters at the grand unified theory (GUT)
 energy scale $M_{GUT}$, {\sl i.e.} $ m_{scalars}(M_{GUT}) \equiv m_0, 
M_{gauginos}(M_{GUT}) \equiv m_{1/2}, A_{trilinear}(M_{GUT}) \equiv A_0$.
Thus, starting from the common scalar soft supersymmetry breaking (SSB) mass

$$
 m_{\tilde{Q}}^2 = m_{\tilde{U}}^2 = m_{\tilde{D}}^2 = m_{\tilde{L}}^2 
= m_{\tilde{E}}^2
= m_{\tilde{H_1}}^2= m_{\tilde{H_2}}^2 \equiv m_0^2,
$$
the common SSB gaugino mass $$ M_1= M_2 = M_3 \equiv m_{1/2},$$
the common SSB trilinear coupling $$ A_t= A_b= A_\tau \equiv A_0$$ and unified
gauge couplings $ \alpha_1 = \alpha_2 = \alpha_3 \equiv \alpha_{GUT}$, all
taken at the GUT scale, the relevant low energy quantities are obtained from
the renormalization group evolution of these parameters from $M_{GUT}$ down
to a scale of the order of the electroweak scale. 
At this scale electroweak
symmetry breaking is required through the minimization equations \cite{mssm},

\begin{eqnarray}
\frac{1}{2} M_Z^2 &=& \frac{\bar{m}_1^2 -  \tan^2(\beta) \bar{m}_2^2 }{ \tan^2 (\beta) - 1}  \label{EWSB1} \\
\sin 2 \beta  &=&  \frac{ 2 B \mu}{\bar{m}_1^2 + \bar{m}_2^2} 
\label{EWSB2}
\end{eqnarray} 

\noindent
where $\bar{m}_i^2 \equiv m_{H_i}^2 + \mu^2 + \mbox{radiative corrections} $,
$\mu$ is the supersymmetric mixing parameter of the two higgs doublets 
superfields and $B$ is the corresponding SSB parameter in the higgs potential.
For a given $\tan \beta$ at the electroweak scale
(defined as the ratio of the two Higgs vacuum expectation values
$<H_2^0>/<H_1^0>$), one determines from the above
equations the $\mu$ parameter (up to a sign ambiguity) and the B parameter
consistent with the physical value of the $Z$ boson mass $M_Z$. In SUSPECT \cite{Suspect} the procedure is carried out including radiative corrections to the 
above EWSB conditions, renormalization group evolution to 1-loop order for the 
soft parameters, and  to 2-loop order for gauge and Yukawa couplings (including
threshold corrections from the supersymmetric spectrum). The full MSSM mass 
spectrum and couplings are computed and fed to DarkSUSY \cite{DSUSY}. 

\noindent
Various phenomenological constraints are taken into account 
(e.g. consistency with top, bottom and $\tau$ masses, present experimental 
limits on superpartner and Higgs masses, limits from bottom decay $ b \to s \gamma$,
no charged LSP, ...), some of which are implemented in SUSPECT and others in DarkSUSY\footnote{We have also disactivated  the $g_\mu - 2$ constraint. 
Clearly a more refined treatment should require consistency with the standard 
model predictions \cite{knecht}. Furthermore, absence of charge and/or color 
breaking minima is not checked within our (exploratory) approach \cite{lemouel} }.\\  

Apart from mSUGRA with independent parameters
$m_0, m_{1/2}, A_0, \tan \beta, sign(\mu)$, we will also consider an alternative
and less constrained case study with 7 free parameters at the
electroweak scale, namely $\mu, M_2, m_A, \tan \beta, m_{\tilde{q}}, A_b, A_t$.
Here $M_2$ is the soft wino mass (we still assume the GUT relation between
$M_2$ and soft bino mass $M_1$), $m_A$ is the physical CP-odd higgs mass, 
$m_{\tilde{q}}$ the common value of {\sl all} soft scalar masses at the electroweak
scale. Furthermore, we do not require
the EWSB constraints Eq.(\ref{EWSB1},\ref{EWSB2}) in this context. 
It should be understood, however, that 
such configurations are considered here only as a test case of the sensitivity of 
gamma fluxes and neutralino relic density to large departure from more physically 
motivated parameterizations such as mSUGRA. We will refer to these configurations as
low energy universality (LEU).\\
Finally, we will also consider configurations motivated by the focus point behaviour
 within mSUGRA \cite{focuspoint}. This allows in principle large $m_0$ values
(in the TeV range) without having to fine-tune the parameters
in Eq.(\ref{EWSB1}). This comes about due to a peculiar behaviour in
the running of $m_{H_2}^2$ from the GUT scale to the electroweak scale,
 which renders this soft mass fairly insensitive to the supersymmetry parameters
provided that $\tan \beta$ is moderate or large.  The implication for neutralino
dark matter can be important as the LSP would acquire a non negligible higgsino
component (in contrast to the generic mSUGRA almost purely bino prediction).
This would affect the relic density as well as the gamma fluxes from 
LSP annihilation into W or Z pairs, when kinematically allowed, are no more suppressed
as compared to the fermion-anti-fermion channels.  Nonetheless, the ``naturalness"
of the focus point is actually moderated by a high sensitivity to the top quark
mass in relic density calculations \cite{Ellisfinetuning}. We will thus adopt
hereafter a qualitative standpoint where $m_0$ is allowed to be in the TeV range,
disregarding fine-tuning issues. We dub these configurations FP-mSUGRA.

\subsection{Relic density}

In addition to the phenomenological or theoretical constraints mentioned 
in the previous
section, one should impose conservatively \cite{siv} that the LSP relic density be in the cosmologically favoured
region $0.1 \lsim \Omega_{\chi} h^{2} \lsim 0.3$. However, we will also
consider in the region $0.025 \lsim \Omega_{\chi} h^{2} \lsim 0.1 $
where we will apply a renormalization procedure on the expected fluxes.

\noindent
Over the last decade the neutralino LSP
relic density has been extensively studied in the literature 
(for reviews cf.~\cite{urb,kamionkowski,bergpr}). 
For simplicity we focus mainly on those regions of the parameter space where
co-annihilation effects \cite{Griest} are unimportant in the estimate of 
$\Omega_{\chi} h^{2}$.
For large $\tan \beta$ values, Higgs resonances can enhance drastically the LSP
annihilation leading to very small $\Omega_{\chi} h^{2}$. Such effects are taken
into account in our study. The actual bottom and top
quark masses become a key issue in this case \cite{roszkowski}. We come
back to this point later on when discussing benchmarks points.

\noindent
When far from co-annihilation and resonance regions, the relic density
can be estimated through \cite{Kolbturner}:
  
\begin{equation}
\Omega_\chi h^2 \simeq \frac{1.07 \times 10^9 x_f}{g_\star^{1/2} M_{pl}(GeV) 
(a + \frac{b}{2 x_f})} \label{freezeout}
\end{equation} 
where $x_f \equiv {m_{\chi}} / {T_f} \sim 25$  provides the approximate 
freeze-out temperature, and $g_{\star}$ counts the massless degrees of
freedom at the typical temperature.  
Here $a$ and $b$ are defined as usual through the Taylor expansion
in relative velocity of the thermally averaged annihilation cross-section
times the neutralino velocity where $<\sigma v>$ \cite{srednicki} is:

\begin{equation}
< \sigma v> = a + b \frac{T}{m_{\chi}} + O(\frac{T^2}{m_{\chi}^{2}}) \label{annrate}
\end{equation}

\noindent
Note that the relation (10) is no more a good
approximation in cases where resonance effects on the 
annihilation rate become important \cite{Griest}.

\subsection{mSUGRA versus general MSSM}

In this section we compare the results of simulations for MSSM (DarkSUSY)
with mSUGRA models (DSS).


\subsubsection{Benchmark points comparisons}
The SUSY benchmark models have been proposed by \cite{ellis,ellis2}
to provide a common way of comparing the SUSY discovery potential
of the future accelerators such as LHC or Linear Colliders. The thirteen SUSY
scenarios correspond to thirteen configurations of the five mSUGRA parameters
with the trilinear coupling parameter $A_0$ set to 0.
The models fulfill the conditions imposed by LEP measurements, 
the $g_{\mu} - 2$ result (which in our case will be not fulfilled), 
and the relic density constraint: $0.1<\Omega_\chi h^2 < 0.3$.

\noindent
The aim of the present section is to derive the gamma ray fluxes for some of
these benchmark models with our current MC simulation programs: DarkSUSY 
\cite{DSUSY} and SUSPECT \cite{Suspect}, which were described previously.
The value of $\Omega_\chi h^2$ is calculated in the DarkSUSY part where
only $\chi^{+}$$\chi^{0}_{2}$ co-annihilations and annihilations related
to the Higgs sector channels are included.
As a consequence of these software limitations, the physically accessible
SUSY domain in this study corresponds to models B,C,G,I and L of 
\cite{ellis,ellis2}.
The simultaneous use of the SUSPECT and DarkSUSY package allows to 
perform RGE evolution from the GUT scale to EWSB scale.
Table \ref{T:models} presents the initial GUT scale parameter values as well as
other input parameters to DarkSUSY, calculated by SUSPECT ($\mu$, $M_1$,
$M_2$, $M_3$, $m_{\tilde q}$, $m_A$). 
The GUT scale and EWSB scale values are also quoted as well as neutralino
mass and gaugino fraction, $R_g$.

\begin{footnotesize}
\begin{table}
\newcommand{\lstrut}{{$\strut\atop\strut$}}
\caption{benchmark models in various simulation.
\label{T:models}}
\vspace{2mm}
\begin{center}
\begin{tabular}{|c||c|c|c|c|c|}
\hline
model & B & C & G & I & L  \\
\hline
$m_{1/2}$  & 255 & 408 & 383  & 358 & 462  \\
$m_0$  & 102 & 93 &  125 & 188 & 326 \\
$tan\beta$  & 10 & 10 & 20 & 35  & 45 \\
sign($\mu$) & + & + & + & + & + \\
\hline
EWSB scale [GeV] & 492.8 & 721.2 & 719.1 & 657.4 & 836.1 \\
GUT scale [GeV] & 2.17\eeen & 0.30\eeen & 1.90\eeen & 1.99\eeen & 1.86\eeen \\
$m_{\chi^0}$ & 99.4 & 162.3 & 155.1 & 144.8 & 190.2 \\
$R_g$ & 0.968 & 0.988  & 0.987 & 0.985 & 0.990 \\
\hline
$\mu$ & 351.5 & 542.5 & 500.0 & 479.2 & 593.7 \\
$m_A$ & 395.8 & 603.9 & 545.4 & 472.4 & 548.4 \\
$M_1$ & 103.6 & 166.0 & 158.5 & 147.7 & 193.1 \\
$M_2$ & 194.7 & 308.0 & 295.4 & 276.0 & 358.6 \\
$M_3$ & 622.7 & 929.2 & 900.3 & 849.6 & 1071. \\
$m_{\tilde q}$ & 537.8 & 787.0 & 767.2 & 718.7 & 907.9 \\
\hline
\end{tabular}
\end{center}
\end{table}
\end{footnotesize}

Table \ref{T:oh2} presents the resulting relic density values with different
theoretical assumptions corresponding either to mSUGRA or to LEU as described
in section 3.1 whereas Table \ref{T:flux} gives results for 
$\gamma$-ray flux, $\Phi_\gamma$
respectively. In order to be free from the Galactic Centre halo modelling,
a renormalisation factor on the astrophysical part of the $\gamma$
flux computation has been applied. This allows a meaningful comparison
between our results and those of {\cite{ellis,ellis2}. 
In case of comparison between mSUGRA and LEU,
one has to choose some common features of the two configurations. We
do this by choosing for the low energy universal masses in the LEU case, 
some of the representative values for the soft masses (and the $\mu$ parameter)
at the electroweak scale, as given by the mSUGRA analysis. In the case
of the soft scalar masses the choice is not unique. The numbers given
in Tables 3 and 4 illustrate typical sensitivities.
 
\begin{footnotesize}
\begin{table}
\newcommand{\lstrut}{{$\strut\atop\strut$}}
\caption{the relic neutralino density, i.e. $\Omega_\chi h^2$, 
as obtained in our simulations assuming a bottom mass of $m_b = 4.61~GeV$
and neglecting radiative corrections.}
\label{T:oh2}
\vspace{2mm}
\begin{center}
\begin{tabular}{|c||c|c|c|c|c|}
\hline
model & B & C & G & I & L  \\
\hline
DarkSUSY    & 1.37 & 4.20 & 1.34 & 0.26 & 0.11 \\
DSS & 0.19 & 0.32 & 0.29 & 0.18 & 0.10 \\
paper [39]  & 0.18 & 0.14 & 0.16 & 0.16 & 0.21 \\
\hline
\end{tabular}
\end{center}
\end{table}
\end{footnotesize}





\begin{footnotesize}
\begin{table}
\newcommand{\lstrut}{{$\strut\atop\strut$}}
\caption{the predicted $\gamma$-ray flux in units of $10^{-12} cm^{-2} s^{-1}$
 from the galactic
center for $\gamma$-rays with energy $E_\gamma > 1~GeV$ 
emanating from M31 within a solid angle $\Theta = 10^{-3}~sr$.}
\label{T:flux}
\vspace{2mm}
\begin{center}
\begin{tabular}{|c||c|c|c|c|c|}
\hline
model & B & C & G & I & L  \\
\hline
G.C. NFW  DarkSUSY           & 293.5 & 29.8 & 207.1 & 1447.0  & 2450.0 \\
G.C. NFW  DSS        & 293.0 & 30.9 & 209.1 & 1442.0  & 2458.2 \\

paper [39]    & 84.29 & 10.19 & 63.90 & 535.0 & 992.4 \\
paper [39] renormalized & 204.8 & 24.76 & 155.3 & 1300 & 2412 \\

M31 NFW   DSS          & 3.5  & 0.37 & 2.5  & 17.3  & 30.0 \\%

 
 
 
\hline

\end{tabular}
\end{center}
\end{table}
\end{footnotesize}

\noindent
As far as benchmark points with large $\tan \beta$ are concerned, the relic
density is very sensitive to the input bottom mass since the latter
controls the position of the s-channel CP-odd Higgs exchange pole
\cite{roszkowski}.

\noindent
The effect of applying radiative corrections has been found to be below 
25\% on flux predictions.
The results obtained here are also compared to those in \cite{ellis}, where a
different mSUGRA MC has been used and a more complete set of co-annihilation
channels has been included.
The observed differences on flux predictions of at most $\sim$ 25$\%$ between 
our results and those in \cite{ellis} may well be explained by different 
treatments of the fragmentation processes with $\pi^{0}$ in the final state.

\noindent
The results given in Table \ref{T:flux} on the $\gamma$-ray flux from the
Galactic Center are given for NFW \cite{nav} halo density 
parameterization, determined above a threshold
of $1$~GeV within a solid angle of $10^{-3}$~sr.
For comparison, the flux from M31 has been also evaluated for 
the same energy threshold and acceptance value, assuming a NFW profile and model 3 in Table 1.


\subsubsection{Predictions from ``wild scan'' simulations}
 
We have also performed a ``wild scan'' by having less restrictive conditions
on the supersymmetric parameters (referred to as "LEU" - see above).
To achieve this, three thousand models have been simulated in the 
$0.025 \lsim \Omega_{\chi} h^{2} \lsim 0.3$ region. 
The value of the lower limit on $\Omega_{\chi} h^{2}$ equal to $0.025$
has been used with a renormalization procedure for flux estimation
described below (for $\Omega_{\chi} h^{2} \lsim 0.1$). The low values
of $\Omega_{\chi} h^{2}$ may indicate that the main component of
the dark matter is not the SUSY LSP and other contributions should be
considered.
The conditions of the performed simulations were for:
\begin{center}
LEU 7 parameters:\\
$10. < \vert {\mu} \vert < 5000.$\\
$10. < \vert M_{2} \vert < 1600.$\\
$10. < m_{A} < 1000$.\\
$1.001 < tan({\beta}) < 60.$\\
$50. < m_{\tilde q} < 1000.$\\
$-3. < A_{t}/m_{\tilde q} < 3.$\\
$-3. < A_{b}/m_{\tilde q} < 3.$\\
\end{center}
\begin{center}
mSUGRA 5 parameters:\\
sign($\mu$) not constraint\\
$50. < m_{0} < 3000.$\\
$50. < m_{1/2} < 1600.$\\
$0.1 < \vert A_{0} \vert  < 2000.$\\
$3. < tan({\beta)} < 60.$\\
\end{center}
Only co-annihilation with $\chi^{+}$ or $\chi^{0}_{2}$ has been considered
which is relevant when the higgsino component of neutralino becomes
substantial. 
All results for the integrated gamma fluxes from M31 as a function 
of the $\chi^{0}_{1}$ mass
were obtained for a NFW profile and model 3 in Table 1, 
and for $\gamma$-ray energy threshold of 30 GeV.
The decay channels with $\pi^{0}$ in the final state subsequently decaying
into $\gamma$s, such as b\={b}, $W^{+}W^{-}$, $Z^{0}Z^{0}$, HH, ..., 
were provided by the PYTHIA 6.1 program which is included in data format
of the DarkSUSY package. 

 

\noindent
Figure 4 shows the integrated flux as a function of the neutralino
relic density $\Omega_\chi h^2$. The shape of the envelope corresponding
to maximal fluxes
can be understood qualitatively by using Eq. (\ref{gr_flux_1}),
Eq. (\ref{freezeout}), and Eq.(\ref{annrate}) and maximizing
$<\sigma v>$ in the parameter space. For instance in the $b \bar{b}$
channel, $<\sigma v>$ reaches a maximum when the lightest sbottom
is degenerate with the LSP leading to $<\sigma v>_{max} \propto m_{\chi}^{-2}$,
that is $\Omega_{\chi}^{min}\propto m_{\chi}^2$. Combining this with the
appropriate dependence of $N_{\gamma}$ on $m_{\chi}$ which can
be obtained from the energy distribution ~\cite{berg7}

\begin{equation}
\frac{dN_\gamma}{dE} = a \frac{e^{-b \frac{E}{m_\chi}}}{m_\chi
(\frac{E}{m_\chi})^{1.5} } \nonumber
\end{equation}
by integrating over the full spectrum,
 
\begin{equation}
\int^\infty_{E_{min}}\frac{dN_\gamma}{dE} dE=
2 a \lbrace \sqrt{\frac{m_\chi}{E_{min}}} e^{-b \frac{E_{min}}{m_\chi}}
+ (Erf[ \sqrt{\frac{b E_{min}}{m_\chi}} ] - 1) \sqrt{b \pi} \rbrace
\end{equation}

\noindent
where $E_{min}$ is an acceptance energy cut, one recovers a behaviour
similar to the envelope shown in Fig 4.

\begin{figure}[!t]
\centerline{\epsfig{file=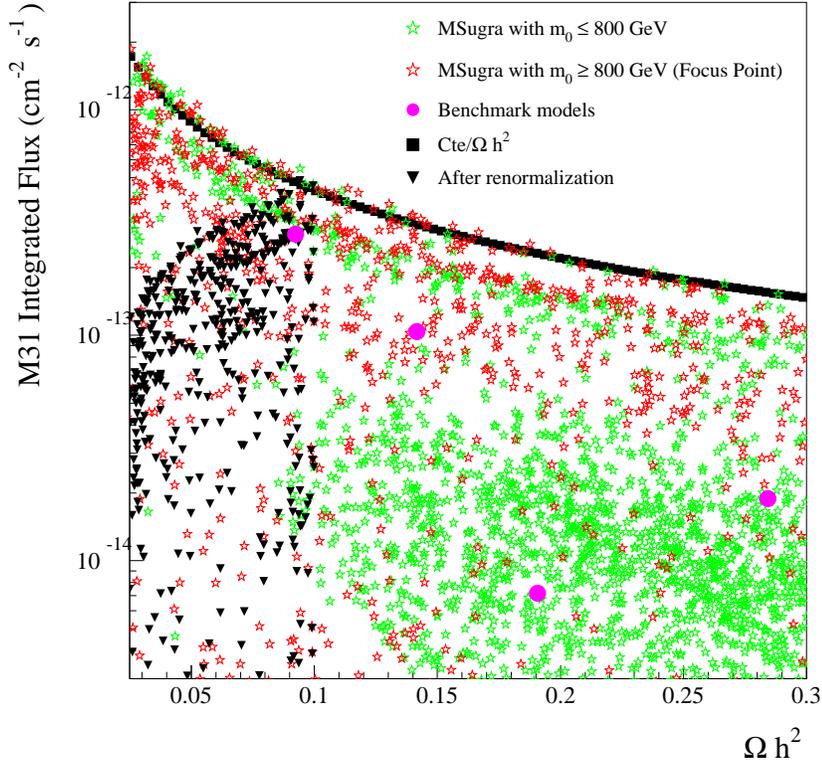,width=12cm}}
\caption{the integrated $\gamma$ flux from M31 as a function of
$\Omega_{\chi}h^{2}$. A renormalization procedure for 
$\Omega_{\chi}h^{2} \leq 0.1$ 
has been applied on flux values as
described in the text.}
\end{figure}

\noindent
A renormalisation procedure has been used for low $\Omega_{\chi}$$h^{2}$ values
where the dark matter halo should not be dominated by the neutralinos.
For values of $\Omega_{\chi}$$h^{2}$ below 0.1 the predicted flux is renormalized
by the ratio $(\Omega_{\chi}$$h^{2}/0.1)^{2}$ to account for neutralinos only
contributing partially to the dark halo. 
As a result of this procedure, the obtained fluxes went down by a factor
of 10 at low values of $\Omega_{\chi}$$h^{2}$. 
The effect of the renormalization procedure described above is also shown in
Fig. 4.

\begin{figure}[!t]
\centerline{\epsfig{file=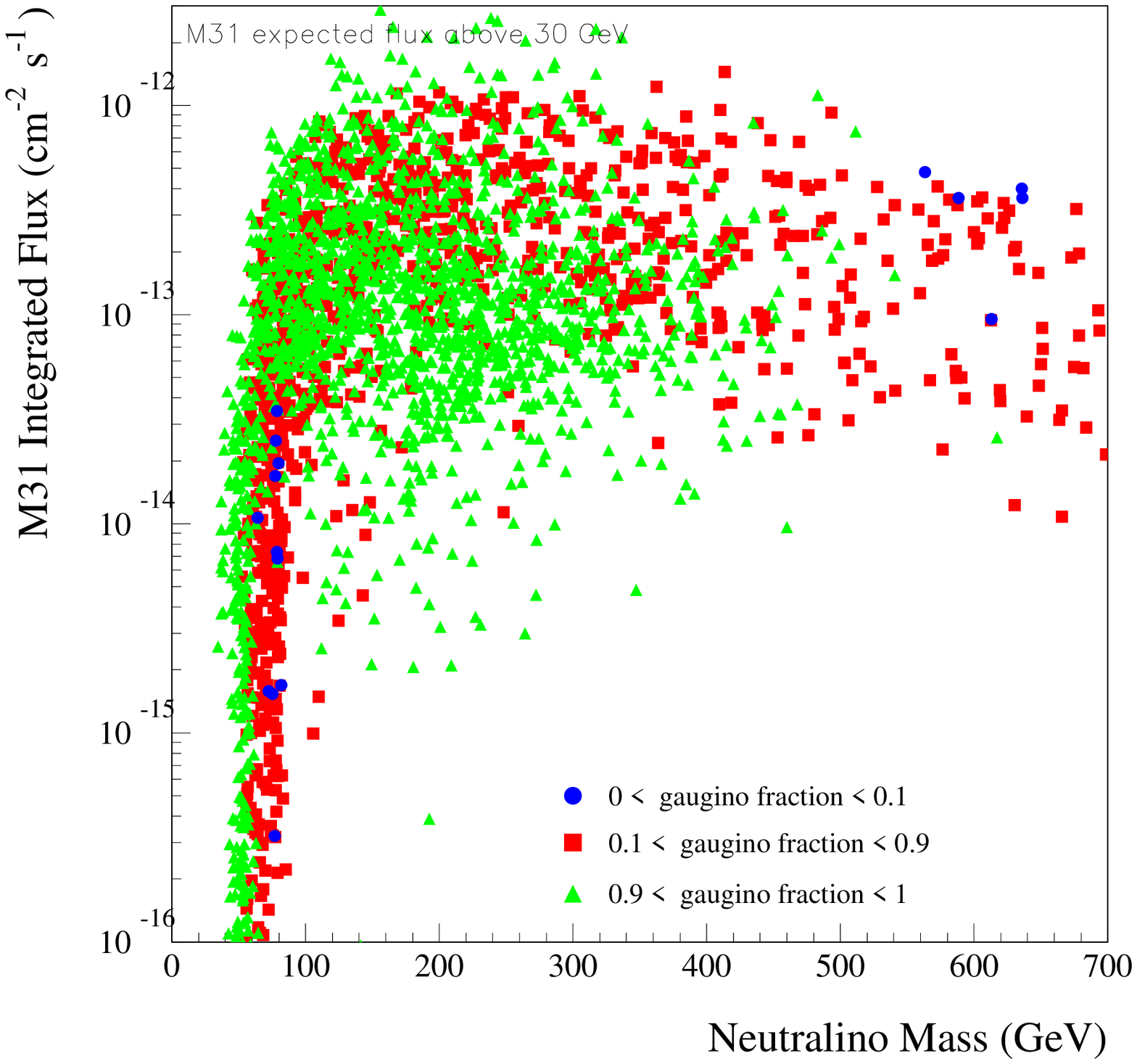,width=12cm}}
\caption{the integrated $\gamma$ flux from M31 as a function of
$m_{\chi}$ for $E_{\gamma} > 30 GeV$. Each point corresponds to a model in
our "wild scan" LEU simulations. 
Three different ranges of gaugino fraction are considered.} 
\end{figure}

\noindent
Figure 5 presents the integrated gamma flux as a function of $m_{\chi}$
for the LEU scheme. The expected fluxes for M31 are of the order of 
$10^{-13}$ $\gamma cm^{-2} s^{-1}$ and show only a weak dependence on 
the gaugino fraction of the
neutralino. Comparing to mSUGRA predictions shown in Fig. 6, clear
differences are observed:\\
- the unification at GUT scale and electroweak symmetry breaking constraints
lead to the exclusion of an important
number of models in mSUGRA models, which are present in the LEU scheme,\\
- the mSUGRA model reduces the number of viable Higgsino like models when
compared to the LEU model, and\\
- mSUGRA models generally show factor $\sim 5$ lower fluxes at low values of 
$m_{\chi}$ as compared to LEU models.\\
The benchmark model points are also presented.

\begin{figure}[!t]
\centerline{\epsfig{file=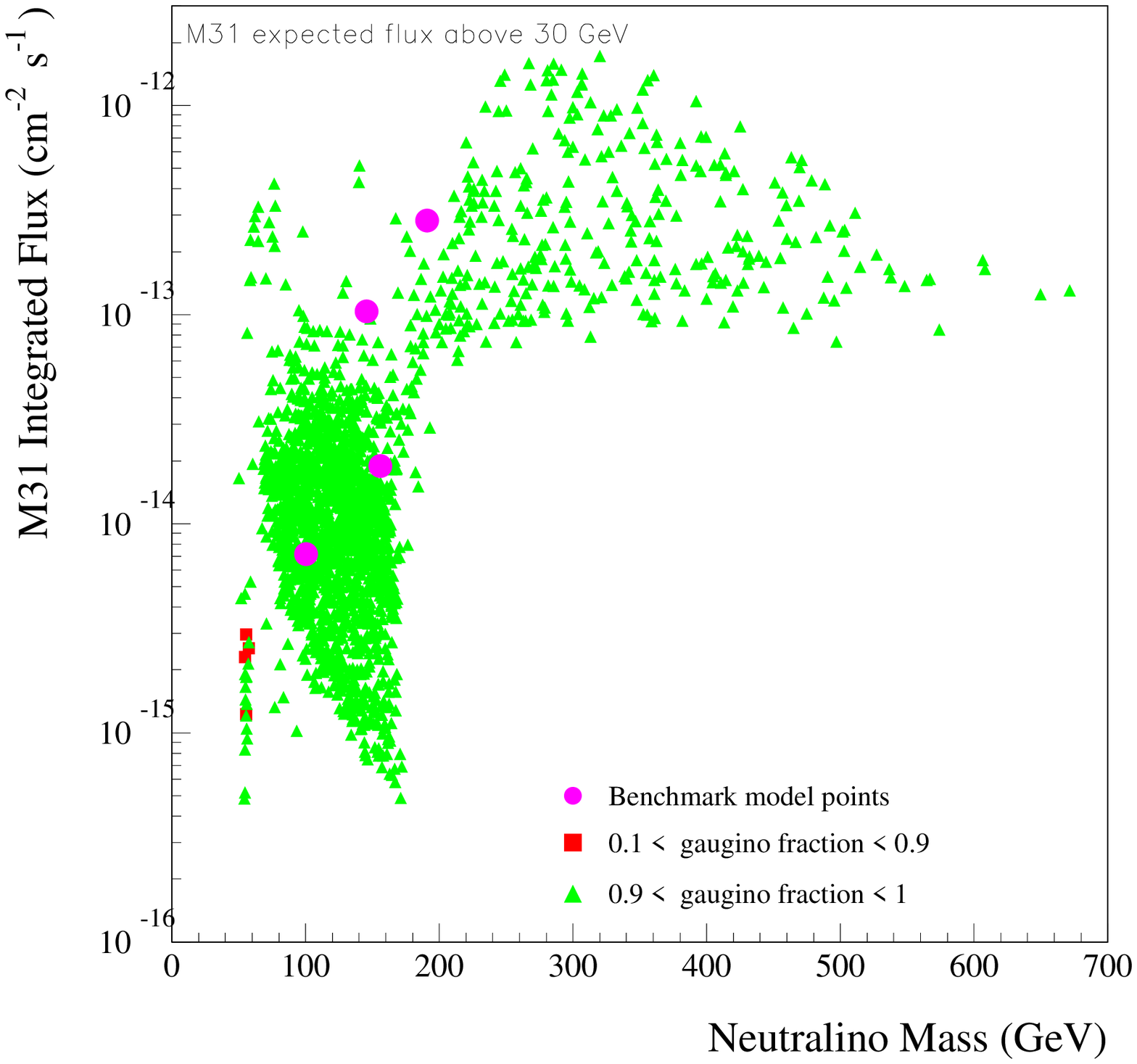,width=12cm}}
\caption{the integrated $\gamma$ flux from M31 as a function of
$m_{\chi}$ in mSUGRA scheme. The constraints applied (as described
in the text) remove large number of LEU allowed configurations. } 
\end{figure}

\noindent
Figure 7 presents mSUGRA results in case of the so-called ``Focus Point''
inspired scenario where $m_{0}$ values may vary up to 3 TeV.
In this case there are more viable models with a neutralino of comparatively
large Higgsino content than in mSUGRA and
the mean fluxes reach values predicted by LEU.

\begin{figure}[!t]
\centerline{\epsfig{file=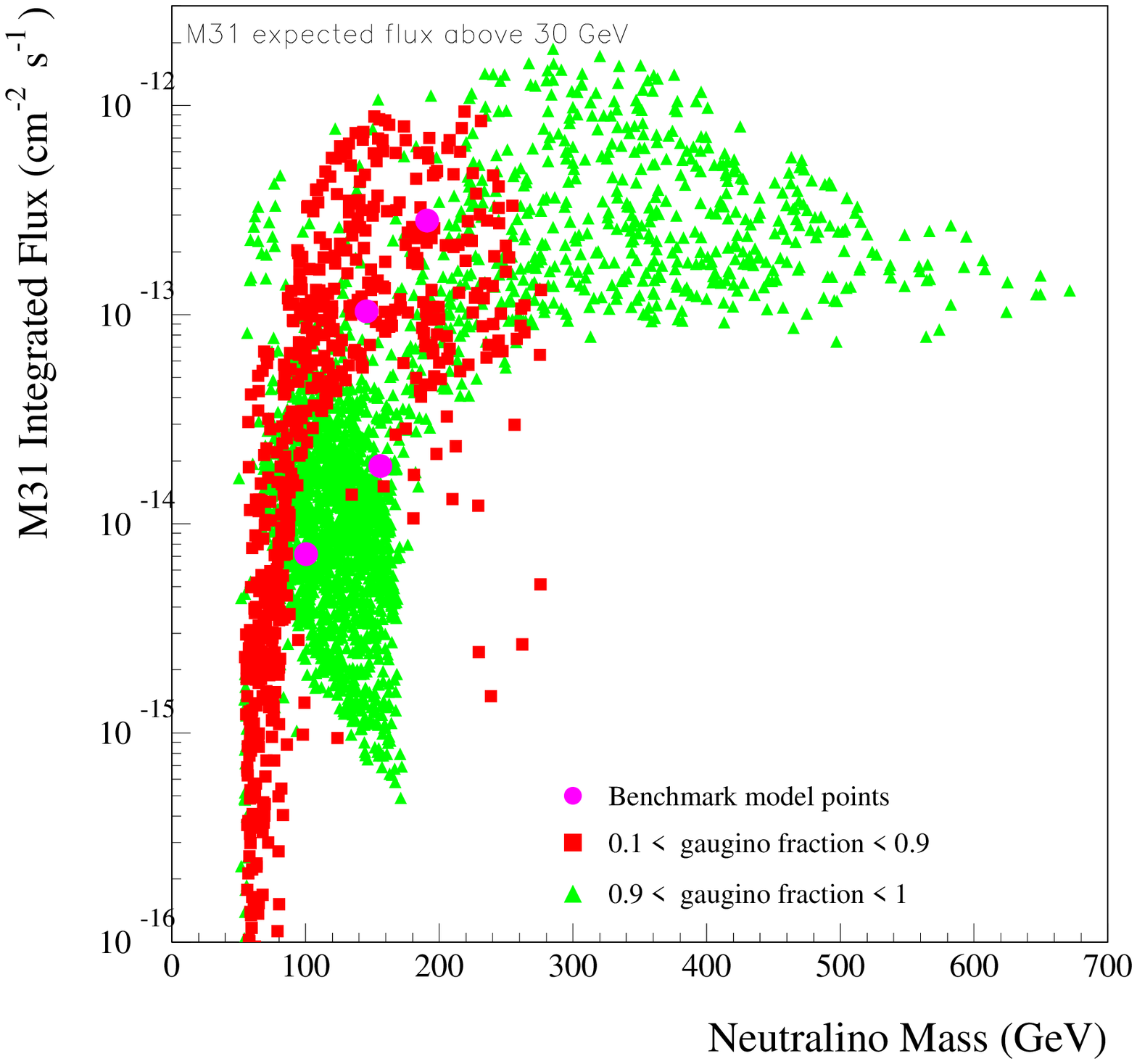,width=12cm}}
\caption{the integrated $\gamma$ flux from M31 as a function of
$m_{\chi}$ in the so-called Focus Point scenario (large $m_{0}$ values).} 
\end{figure}

\noindent
The CP-odd Higgs mass resonance effect (i.e. $m_A\approx 2\,m_{\chi^0}$) on the
annihilation cross section and predicted fluxes has also been examined.
Figure 8 presents an increase of the
integrated flux at the $m_{A}$ pole by at least an order of magnitude
compared to the region far from the resonance.
This effect reflects increase of the predicted fluxes for large $tan\beta$
shown in Fig. 9, as the CP-odd Higgs s-channel dominates at 
large $tan{\beta}$ in the mSUGRA frame.

\begin{figure}[!t]
\centerline{\epsfig{file=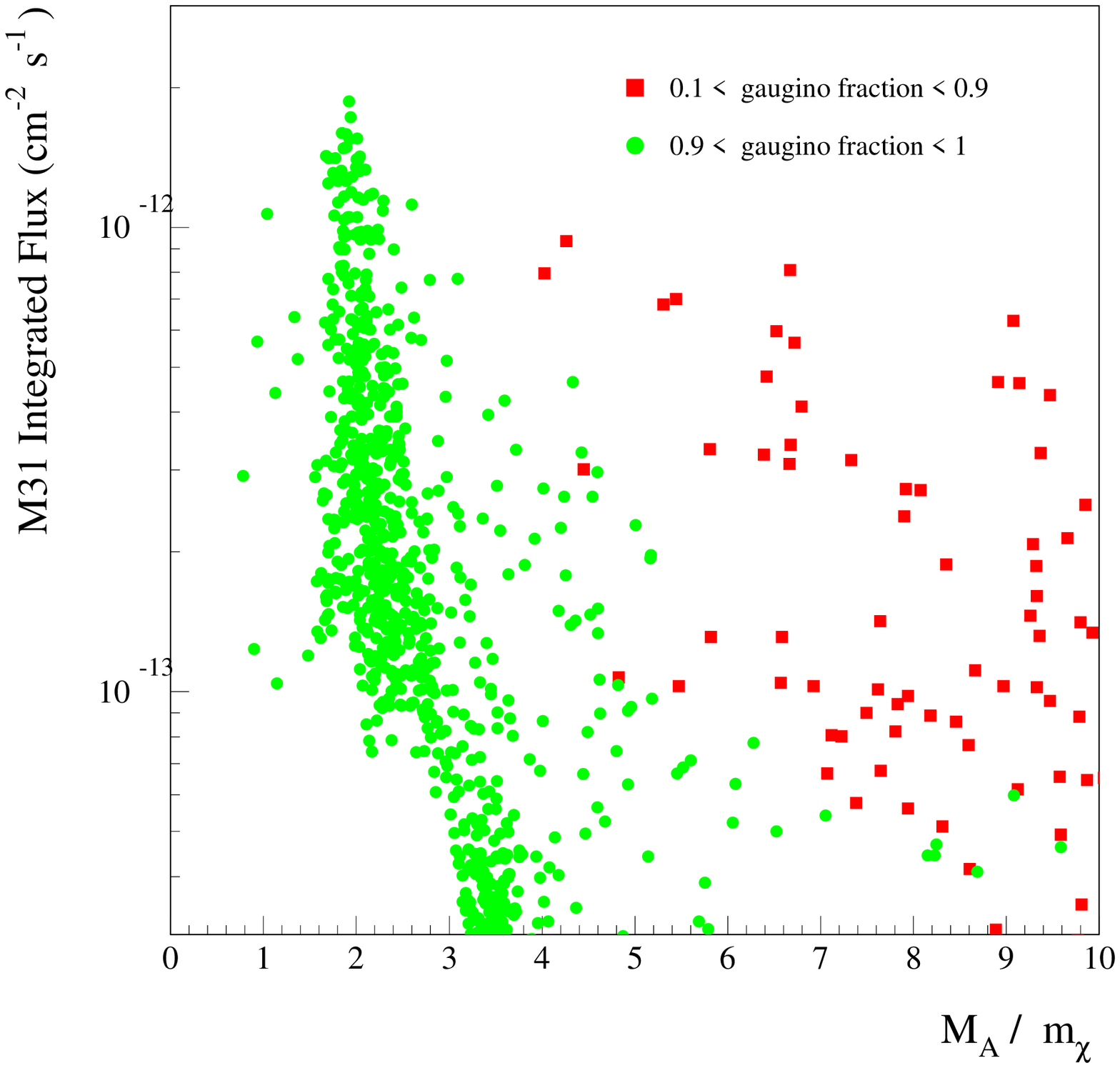,width=12cm}}
\caption{the integrated $\gamma$ flux from M31 as a function of
$M_{A} / m_{\chi}$. Here only fluxes above $10^{-14}$ are presented. 
A large enhancement at $2 m_{\chi}$ equal to $M_{A}$ is observed.} 
\end{figure}

\begin{figure}[!t]
\centerline{\epsfig{file=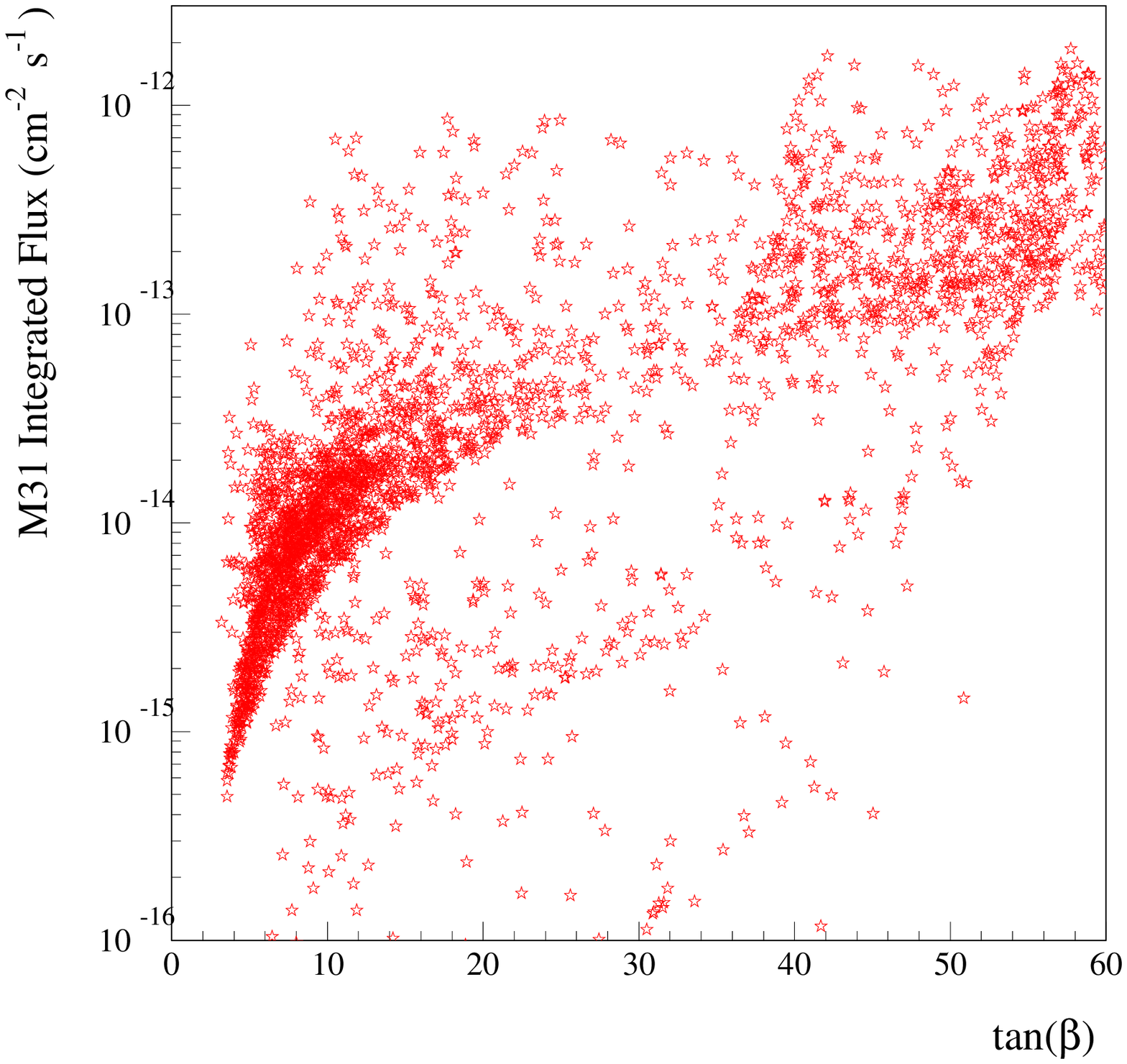,width=12cm}}
\caption{the integrated $\gamma$ flux from M31 as a function of
$tan{\beta}$. Here the highest fluxes also correspond to the CP-odd
Higgs contributions. } 
\end{figure}

\section{Detection of Dark Matter in M31 with\\
         CELESTE}

\subsection{The CELESTE experiment}

The ``Cherenkov Low Energy Sampling and Timing Experiment'' (CELESTE) is located on the site of Th\'emis
in the French Pyr\'en\'ees. It detects $\gamma$-rays in the high energy domain and was designed to fill
the energy gap between satellites and imaging atmospheric Cherenkov telescopes (IACTs). While the former have
been so far limited to $\sim$$10\:$GeV because of a too small effective detection area, the energy domain covered
by the latter starts at $\sim$$200\:$GeV due to the noise arising from the night sky background.
The energy window around $50\:$GeV has been opened by CELESTE during the winter 1999-2000 by the detection of the
Crab nebula and its flux measurement at $60\:$GeV~\cite{deNaurois2002}. Nearly at the same time, CELESTE detected
the Active Galactic Nuclei (AGN) Mkn~421 during a series of flares.
 It was the first sub-$100\:$GeV detection of
this nearby blazar and the emission observed by CELESTE showed a clear correlation with the flux recorded above
$250\:$GeV by the CAT telescope, operating on the same site~\cite{Holder2001}.
Before these results, the sub-$100\:$GeV region was the only part of the electromagnetic spectrum which remained unexplored.
The access to this region is of great importance for a number of key questions in high-energy gamma-ray astrophysics.
For instance, the high-energy part of some AGNs can dominate the 
 entire spectrum,
giving new insight into their structure as well as the particle acceleration and 
cooling processes occuring in their vicinity.
Pulsars could be also very promising sources : since their $\gamma$-ray spectrum is 
believed to decrease sharply at a few tens
of GeV, the observation of the predicted spectral cutoff should strongly constrain 
the models describing the
particle acceleration in their magnetosphere. Concerning galactic cosmic-rays, the current belief of their acceleration in
shell-like supernov\ae~remnants could be also tested in a few cases by the observation of high-energy
$\gamma$-rays.
Apart from astrophysical issues, we propose here to use CELESTE for the study of 
dark matter and searches for the indirect signatures of
neutralinos, namely their gamma-ray signal
above $50\:$GeV coming from neutralino annihilations in M31. At the 
latitude of Th\'emis ($42^\circ$ North), M31
transits very close to Zenith.\\

\noindent
The CELESTE experimental setup is fully described in~\cite{deNaurois2002}. This detector records the Cherenkov light emitted
by the secondary particles produced during the development of the cosmic-ray atmospheric showers. In order to reduce the night-sky noise
and to achieve a low energy threshold, CELESTE uses $40$ heliostats ($53$ since January 2002) of the former solar plant in Th\'emis, with
a total collection area of $\sim$$2000\:$m$^2$. An efficient discrimination between $\gamma$ and hadron-induced showers allows to
extract the $\gamma$-ray signal from the more abundant charged cosmic-ray background (mostly protons, helium nuclei and electrons). 
The event analysis is based on the differences between
both types of showers regarding the homogeneity and the time dispersion of the 
Cherenkov light pool as sampled by the heliostats.
After the detector trigger and all analysis cuts, the $\gamma$-ray acceptance at Zenith has been parameterized as a function of the
energy~\cite{LeGallou} :

\begin{equation}
\mathcal{A}(E)=A_0\bigg[1-\exp\bigg(-\frac{E-E_0}{E_c}\bigg)\bigg],
\label{celeste_acc}
\end{equation}
with $A_0=1.37\times10^4\:$m$^2$, $E_0=34.7\:$GeV and $E_c=56.2\:$GeV.

\subsection{Simulation results}

The flux predictions of mSUGRA models 
were folded with the acceptance function of the CELESTE detector.
Figure 10 shows the expected count rates from M31 in units of $\gamma$/min.
The most optimistic models predicts values of the order of
$10^{-3}$, which should be compared with the $4.9\:\gamma$/min which are observed
by CELESTE from the Crab nebula (the standard candle for high-energy $\gamma$-ray astronomy)~\cite{LeGallou}.
Note that these results were obtained without any signal enhancement
effect expected from clumpiness of the dark matter halo or possible black hole
neutralino accretion. Since the sensitivity level of atmospheric Cherenkov detectors is proportional
to $1/\sqrt{T_\mathrm{obs}}$, only sources with a flux of a few tenths
of the Crab nebula flux can be detected within realistic observation
times. Therefore the overall enhancement due to astrophysical effects
has to exceed at least a factor of 100 for detection of a $\gamma$-ray signal
from neutralino annihilations to become feasible. 
The impact of such effects will
be addressed in the next section.\\

\begin{figure}[!t]
\centerline{\epsfig{file=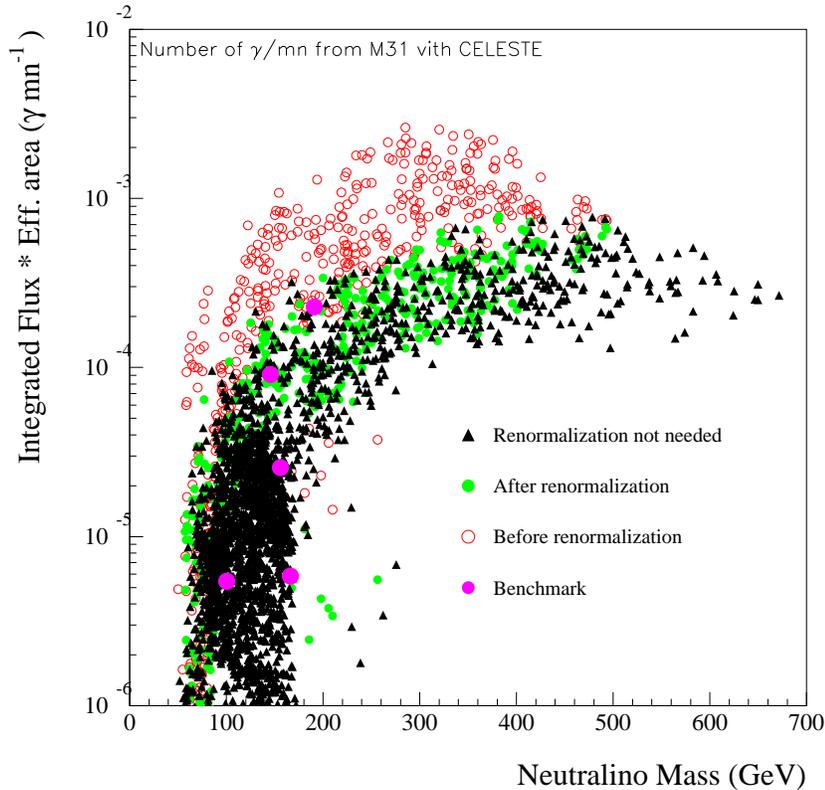,width=12cm}}
\caption{number of $\gamma / min$ as expexted from M31 with CELESTE
acceptance in mSUGRA models. } 
\end{figure}

\noindent
A possibility of the shape discrimination with the energy differential 
flux measurement in CELESTE has been also studied. The expected
shape for $\gamma$-rays produced in $\chi^{0}\chi^{0}$ annihilations
are described by a decreasing exponential function with the exponent 
strongly depending on $m_{\chi}$. This sharp exponential cut-off
should be compared with standard astrophysical source power law
expectations. 

\begin{figure}[!t]
\centerline{\epsfig{file=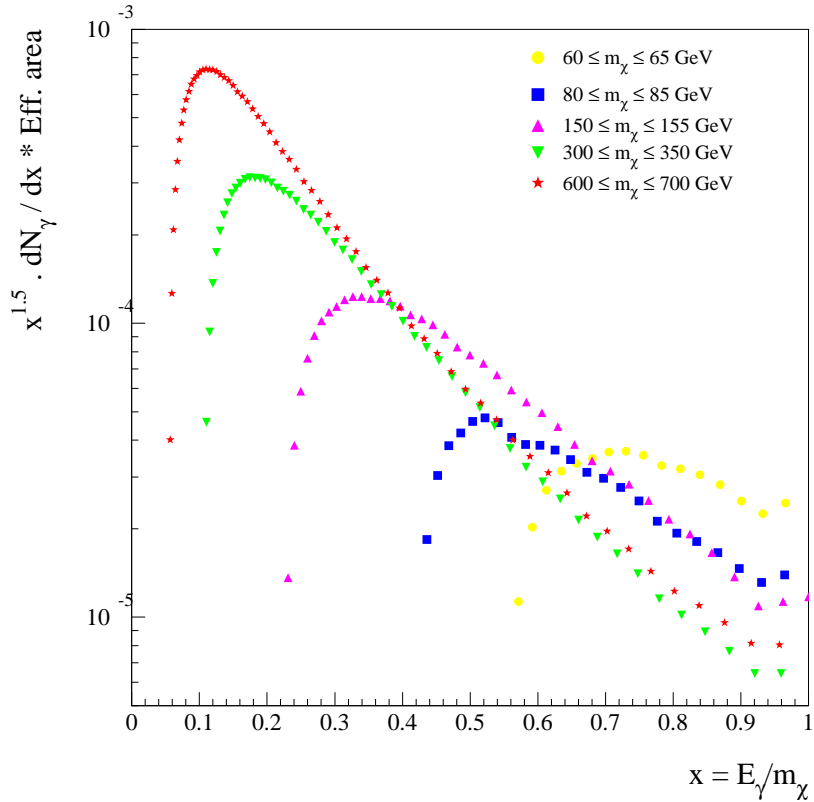,width=12cm}}
\caption{a shape variation of the scaling variable x equal to
 $E_{\gamma} / m_{\chi}$ \cite{berg7} for various $m_{\chi}$ values. 
A weak dependence on $m_{\chi}$ comes from different annihilation
channels.  } 
\end{figure}

\noindent
The shape discrimination relies on the assumption of the $\gamma$
detection from the considered source, in our case M31.
The event selection and cuts on the kinematical variables of the
$\gamma$ candidates allow to enhance the $\gamma$ contribution
to the measured sample with respect to the charged cosmic-ray initiated 
showers. In order to reduce further the charged
cosmic-ray background and to optimize the signal to background ratio, various
procedures such as a sliding window type method can be considered. This
analysis will be addressed in details elsewhere.

\noindent
Figure 11 presents predicted energy differential distributions
for various $m_{\chi}$ ranges after the renormalization of the
total integrated fluxes. Here an almost scaling variable equal to 
$E_{\gamma}/ m_{\chi}$ was used as proposed by \cite{berg7}. 
No background contribution has been taken
into account here. These results suggest an increase in
sensitivity with $m_{\chi}$. However, due to the CELESTE acceptance 
varying with energy, a realistic discrimination measurements can only 
be considered for neutralino masses above 200 GeV. 
It is interesting to note that only models with substantial co-annihilation
and large $tan{\beta}$ can lead to acceptable dark matter densities for 
${\Omega_\chi}$$h^{2}$ above 200 GeV.

\section{Impact of Astrophysical parameters}

\subsection{Halo Clumpiness}

\medskip
Early simulations of the growth of CDM halos found smooth tri-axial 
structures, but higher resolution simulations revealed that
a significant fraction of the mass may lie in substructures (sub-halos)
orbiting in virialized dark matter halos ~\cite{bb14,bb15,bb16}.
Such substructures seem to extend down to the smallest mass scale, 
possibly well below that of the most massive globular clusters.

\medskip

\noindent
The CDM scenario, however, predicts a larger number of substructures 
in galactic halos than the observed number of dwarf satellites in
the Milky Way~\cite{bb15,bb40}. The cases of gas expulsion by supernova-driven 
winds or 
the failure of gathering enough gas for stellar formation were
considered in ~\cite{bb15} as possible explanations. In this case there would be a large number 
of clumps with few stars, and some of the known high-velocity gas clouds may 
possibly trace them.
It has been also shown recently that there is good agreement
between the rotation curves of the largest satellites and those found
in N-body simulations, provided star formation is less efficient in
these satellites ~\cite{swts} .
The possibility that the anomalous image flux ratios 
observed in several gravitational lenses could be explained by low mass 
satellites in 
the lensing galaxy was also explored ~\cite{bb42}. It was found that the 
required 
projected satellite mass fraction is of the same order as those predicted in
CDM simulations,
illustrating that there is general agreement
between the underlying theory for structure formation and observed 
structures. 

\noindent
In order to estimate the effect of clumpiness we calculate first the
contribution to the flux of a DRACO-type object with 
virial velocity $v = 10~\rm km~s^{-1}$, and central mass 
$M \approx 10^8~M_\odot$
modeled by using a NFW profile
with scale radius $r_s = 0.4~\rm kpc$ and density $\rho_0 =
6 \times 10^{-24}~\rm g~ cm^{-3}$. This is obtained by calculating the
sum of the line of sight integral $\Sigma$, as defined in Sect. 2,
over the field of view  of CELESTE which we denote by ${\Sigma}^{10}$
 (${\Sigma}^{10} = {\Sigma}^{19}(3.5 kpc)$ defined in section 2) .
It is expressed in units of $\rm \equiv 10^{19}~GeV^2~cm^{-5}$ as before.
Such a single DRACO-type clump orbiting around M31 would yield 
${\Sigma_{\rm D}}^{10} \approx 0.03$, to be compared with
${\Sigma}_{\rm M 31}^{10} \approx 3$ for M31 (cf. Sec. 2). 

\noindent
\medskip
Now we  consider the mass and the radial distribution of clumps orbiting 
around a halo. The sub-halo mass function is a power law close to
$dn(m)/dm \propto (m/M_H)^{-1.9}$ ~\cite{bb14} where $M_H$ is the mass of 
the halo.
The radial distribution of clumps in a halo, deduced from N-body simulations,
behaves as $n(r) \propto \left[ 1 + \left(r/r_c \right)^2 \right]^{-3/2}$ 
~\cite{bb31}.
Combining these two equations we obtain
the distribution of clumps having a mass $m$ at radial distance $r$: 
\begin{equation}
 n(r, m) \propto (m/M_H)^{-1.9} \left[ 1 + \left(r/r_c \right)^2 \right]^{-3/2}
\propto v^{-3.8} \left[ 1 + \left(r/r_c \right)^2 \right]^{-3/2}
\end{equation}
where $v$ is the virial velocity of the clump.
This equation is normalized to the number of clumps of a given
mass predicted by CDM simulations.

\noindent
About 1000 clumps with velocities above 
$v = 10~\rm km~s^{-1}$ are expected
in a galactic halo having the mass of M31 ~\cite{bb6}.
We truncate the clump mass distribution between $ 1 \leq v \leq 10~\rm km~s^{-1}$.
We consider two spatial distributions of clumps: a compact distribution 
with core radius  $r_c \sim 16~\rm kpc$ and a more extended distribution with
$r_c \sim 30~\rm kpc$. Both are independent of the clump mass.
In these profiles, around $ 75 \%$ of the clumps are within the core radius.

\medskip
\noindent
The total integrated contribution of clumps with $v \geq 1~\rm km s^{-1}$ is
${\Sigma} \approx 2500 \times {\Sigma}_{\rm D} = 77$. 
Therefore the flux from M31 may be  boosted by a clumpiness factor $\sim 77/3 \approx 25$.
The  clumps in 
the field of view of CELESTE yield respectively ${\Sigma}^{10} = 
20$ for 
$r_c \sim 16~\rm kpc$ and ${\Sigma}^{10} = 10$ for
$r_c \sim 30~\rm kpc$, that is a factor $3.3 - 6.6$ larger than
the smooth profile of M31.

\medskip

\noindent
Globular clusters are known to continuously loose stars due to tidal forces 
from the halo gravitational potential. Mass is stripped from the outer 
regions of globular clusters and evaporates mainly at each passage through 
the disk. This is why it is generally argued that globular clusters do not 
have dark matter except perhaps in an inner bubble.
The structural evolution of sub-halos was recently analyzed in ~\cite{bb5},
which shows that the tidal radius of distant satellites is well beyond 
their optical radius. 
This may also explain why the distant globular cluster Palomar 13 
(at a distance of $24.3~\rm kpc$) is well fitted by a NFW profile ~\cite{bb18}, i.e. 
this anomalous globular 
cluster could, in fact, be a Milky Way dark clump
which has not been destroyed or modified by the galactic tidal field ~\cite{bb19}. 
The central mass-to-light ratio of the distant satellites Palomar 13 and DRACO 
(at a distance of 80~kpc) are, respectively, $40~(M/L)_\odot$ and 
$80~(M/L)_\odot$, whereas for 
nearby globular clusters it is 
approximately $3~(M/L)_\odot$. We assume that this is the signature of enhanced particle
evaporation in the inner region of clumps. To account empirically for this 
effect we have divided by 2 the central density of the clumps at radial 
distance $r = 20~\rm kpc$, by 20 that of objects at $r = 10~\rm kpc$ and 
have ignored clumps at smaller radius. With this hypothesis the most massive
objects within $r = 10~\rm kpc$ have a mass $< 5 \times 10^6~M_\odot$. In any
case, a significant fraction of more massive objects is forbidden by the
observations because it would dynamically heat the disk (see below).

\medskip 
\noindent
With the hypothesis of evaporation, the integrated 
contribution of clumps is ${\Sigma} = 11 - 34$ for the whole of M31 and 
to ${\Sigma}^{10} = 0.75 - 2.2 $ in CELESTE's 
field of view, indicating only very modest enhancement factors of the $\gamma$-ray
signal. 

\medskip
\noindent
If a large number of massive satellites orbit within the M31 halo,
their passages through the disk causes disk heating ~\cite{bb27} or may induce
warping motions ~\cite{bb28}. This is confirmed by recent N-body CDM 
simulations of the Milky Way ~\cite{bb30}. In the case of 
a $\rm \Lambda CDM$ cosmology, however, 
the clumps are less massive and are located at larger radial distance than 
in standard CDM, so only a small number get near the disk. In that case 
they do not heat the disk efficiently ~\cite{bb29,bb30}. We have already 
approximately accounted 
for the first effect by ignoring the most massive clumps with mass 
$10^9~M_\odot$ in the inner regions of galaxy. In order to account 
for an ``extended halo distribution'', as in ~\cite{bb30}, we have considered 
the 
same clump distribution as before but with core radius $\rm 100~kpc$ and 
the same mass spectrum. With that distribution the clump flux in CELESTE's
field of view is ${\Sigma}^{10} = 2.2$ without evaporation, 
or ${\Sigma}^{10} = 1.0$ with the hypothesis of evaporation.
 
\medskip

\noindent 
These $\gamma$ ray signal enhancement factors due to halo clumpiness
should be compared to those determined by N-body simulations of Milky Way
type objects. Whereas earlier results by one group ~\cite{cal} indicate
a two order of magnitude effect, more recent simulations ~\cite{sto} 
find only a factor $\sim$ 2 enhancement. Another study performed  
on the clumpiness enhancement
factor, in particular in case of M87, observation ~\cite{baltz2} 
predict values between 13 and 40.  
These calculations were extended for Milky Way or M31 type galaxy case leading
 to the values of the clumpiness enhancement factor of the order of 
20 ~\cite{pier}.  

\subsection{Black hole accretion}

\medskip
The central region of M31 contains a very compact object
with a mass $\sim 3.6 \times 10^7~M_\odot$ which is assumed to be
a supermassive black hole (SBH) ~\cite{bb32,bb33}. The dynamical effect of
a SBH on a distribution of nearby WIMPS was studied 
by ~\cite{bb33} in the case of the Milky Way. The adiabatic growth of a 
SBH at the
center of a halo produces a spike with density $\rho \propto r^{- \gamma}$,
$2.2 \leq \gamma \leq 2.5$ within a radius $\rm \sim 10~pc$ for a SBH
like that in M31. In such a spike the density reaches the annihilation 
density $\rm 10^8~M_\odot~pc^{-3}$ in a significant region surrounding 
the black hole producing a huge enhancement of the annihilation flux. 
This calculation assumes a very high accretion
rate and a stable dynamical regime during an extremely long period. 
In fact the accretion rate for the assumed black hole adiabatic growth is much higher
than the current estimates of present-day growth-rates ~\cite{bb37, bb38}.
Several authors have discussed more realistic scenarios (i.e.~\cite{bb34, 
bb35}). 
In the CDM scenario, large halos grow through the merging of smaller
building blocks. Galaxies as massive as the Milky Way or M31 have certainly 
experienced significant merging in the last Gyrs. There may be several
SBH progenitors coming from the mergers, a SBH may be spiraling into
the galactic potential. Motions of the SBH transfer energy to the particles
lowering their density ~\cite{bb36}. Thus, even if at early times
a massive BH was present in a dense environment resulting in a rapid
adiabatic growth of the BH surrounded by a dense spike, there are a
number of accidental events and perturbations during the life of
the SBH which all tend to destroy the spike. Nevertheless a SBH
will continuously accrete material. If it is coinciding with the dynamic 
center of a galaxy for a sufficiently long period it will built
a central dark matter density cusp. 

\noindent
We performed
N-body simulations assuming initial dark matter density profiles of
$\rho \propto r^{-\alpha} (1 + r/r_c)^{-\beta}$ with $0 \leq \alpha \leq 1$
and under the hypothesis of the existence of a massive central black 
hole - SBH.
These simulations were performed with the aid of the
public code Gadget ~\cite{bb39}. In each case, the density 
distribution reaches a stable limit with $\rho \propto r^{-1.5}$, 
independent of the initial profile, within dynamical time of $10^{6}$ yrs. 
An example of such simulations is shown 
in Fig. 12. This illustrates that a stable configuration 
which may be build in within a relatively short time. 


\begin{figure}[!t]
\centerline{\epsfig{file=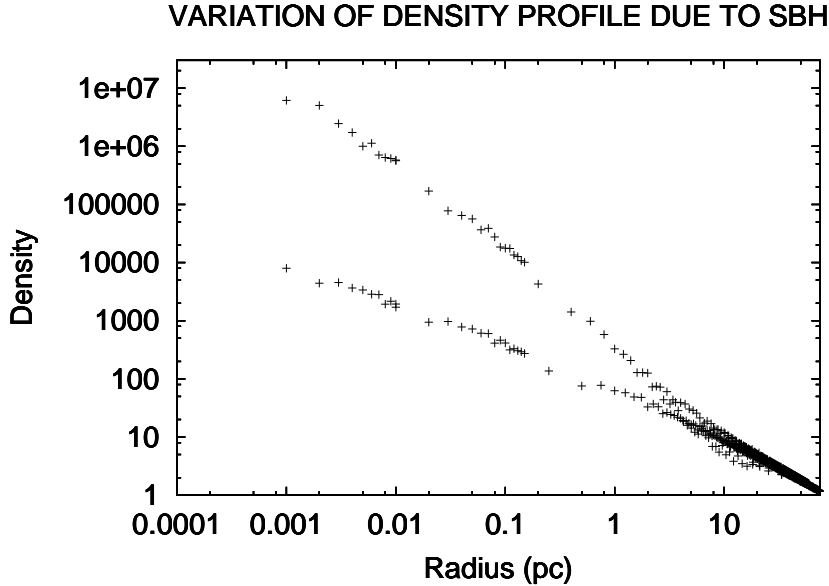}}
\caption{
example of simulation showing the variation of density profile
due to a massive SBH of $3.7 \times 10^7~M_\odot$. The initial profile is 
in $\rho \propto r^{-0.75}$. The final profile in $\rho \propto r^{-1.5}$ 
is reached in less than $10^6$ yrs. The density is in $M_\odot~\rm pc^{-3}$.}
\label{figure_SBHprof}
\end{figure}

\noindent
The $\gamma$-ray flux enhancement due to the existence of such a dark matter
spike as compared to a NFW profile is of the order
of $1 + [ln (r_{max} / r_{min}) -1]~ r_{max}/R$
where $r_{min}$ and $r_{max}$ are respectively the minimun and maximum radius
of the profile in $r^{-1.5}$ and $R$ is the radius of the initial region
with profile in $r^{-1}$. Taking $R = 3.5~\rm kpc$, 
$20 \leq r_{max} \leq 100~\rm pc$, and $r_{min} = 10^{-5}~\rm pc$ which
is $\sim 11$ Schwarzschild radii, we get an only modest enhancement factor between
1.08 and 1.45. It is seen that the enhancement factor is
small, nevertheless, the presence of the SBH in this complicated multicomponent
object warrants a singular profile at small radius. This is also
important with regard to evidence that 
the universal singular CDM profiles ~\cite{aa14, aa19} are not 
seen in the dark halos of low surface brightness galaxies 
~\cite{aa6}. In these objects, however, there are numerous suggestions for 
a hidden baryonic component ~\cite{aa3, aa17, aa10} which may modify its 
inner profile complicating 
the interpretation of dwarf halos ~\cite{aa7}.  

\noindent
Finally in some cases the hypothesis of a central spike may be valid.
This may happen if there is a density increase during a short period
near the SBH, for example when a massive clump is falling in the
SBH. As soon as the profile is steeper than $r^{-1.5}$ the logarithmic
term in the enhancement factor is replaced by a power of the inner radius.
In the case of a singular isothermal (SIS) spike of $1~\rm pc$ the total
flux within $3500~\rm pc$ is boosted by a factor 29.  
The impact of clumpiness and black holes are summarized in 
Tab.~\ref{table_prof}. The global enhancement factor on expected flux
may vary between 5 and 100, depending on the clumpiness of the halo
and CDM-accreting SBH. In particular, the large variation between 
${\Sigma}$ and ${\Sigma}^{10}$ in case of an extended clump distribution
(last line of Table 5) is due to the small angular size of the instrument
compared to the compact distribution case (line 3 of Table 5). 

\begin{table}
\begin{center}
\begin{tabular}{|c c c c c|}
\hline\hline
Clump distribution & Dynamics & SBH environment & $\Sigma$  & 
$\Sigma^{10}$ \\
\hline
 compact & no evaporation & SIS spike & 165 & 100 \\
 extended & no evaporation & SIS spike &  140 & 90 \\
 compact & no evaporation & $\rho \propto r^{-1.5}$ & 80 & 20 \\
 extended & no evaporation  & $\rho \propto r^{-1.5}$ &  55 & 6 \\
 compact & evaporation & $\rho \propto r^{-1.5}$ & 25 & 6  \\
 extended & evaporation  & $\rho \propto r^{-1.5}$ & 40 & 5 \\
\hline\hline
\end{tabular}
\label{table_prof}
\caption{impact of astrophysical parameters such as clumpiness of
the M31 halo and SBH in its centre,on flux predictions, 
smooth and clump contributions added.}
\end{center}
\end{table}

\section*{Conclusions}

The putative dark matter content of M31 has been estimated and
modelled from its observed rotation velocity profile, by adding a NFW
dark halo to the observed disk and bulge component.

\noindent
The study of the benchmark models in case of 
M31 allowed to qualify the DarkSUSY and SUSPECT MC programs used
for gamma flux predictions in CELESTE experiment. Only models I and L
(large $tan\beta$) could be considered as favourable for the detection.
This conclusion is confirmed by a further study of the so-called ``wild
scan'' simulation in mSUGRA scheme. Various aspects such as dependence
of the results on heavy quark masses or CP-odd higgs pole contribution
have been investigated.

\noindent
We have taken into account the atmospheric electron and CR proton
backgrounds to the M31 signal and applied the CELESTE gamma--ray
energy dependent acceptance factors. Assuming a smooth neutralino
halo around M31 leads to photon rates at the telescope of order
a few $10^{-3}$ counts per minute. The strongest signal corresponds
to supersymmetric configurations where the neutralino mass $m_{\chi}$
exceeds $\sim$ 200 GeV.
As a realisitic observation would require at least a signal of order
a tenth of that of the CRAB -- 4.9 photons per minute -- we
estimate that a gamma--ray annihilation signal from M31 is beyond
the reach of an atmospheric Cerenkov detector of the CELESTE
generation if the neutralinos are smoothly distributed in that galaxy.
On the other hand, if the halo is made of clumps with inner profiles
\`a la Moore or if there is strong accretion going on around the central
black hole, the expected signal may be enhanced by two orders of
magnitude and could become detectable. We conclude therefore that a
survey of M31 with CELESTE is worth being undertaken.


\section*{Acknowledgements}
We would like to thank Jean-Loic Kneur for his help in running the
SUSPECT program and building the interface between SUSPECT and
DarkSUSY.
We are also very much indebted to Piero Ullio for the time he spent
on useful discussions about the DarkSUSY program and the physics related
to the studied subjects.
We thank the French GDR ``PCHE'' groups for supporting this study. 

\newpage




\end{document}